\documentclass[12pt,a4paper]{article}
 \usepackage{jheppub}

 \title{Quantising Chiral Bosons  On Riemann Surfaces}

\abstract{Sen's action in two dimensions governs a chiral boson coupled to a two-dimensional metric together with a second chiral boson that couples to a flat two-dimensional metric. This second scalar  decouples from the physical degrees of freedom. The generalisation of this action to one in which the second chiral scalar couples to an arbitrary second metric is used to formulate the theory on an arbitrary two-dimensional manifold. We use this action with both metrics Riemannian (or complex) to formulate the path integral on any Riemann surface. We calculate the partition function in this way and check the result with that   calculated using  canonical quantisation, and then extend this to multiple chiral bosons.
The partition function for chiral scalars taking values on a rational torus
is a sum of terms, each of which is the product of   two holomorphic functions, one a   function of the modulus of the first metric and the other a   function of the modulus of the second metric.
In particular, for the case of chiral bosons moving on a torus defined by   an even self-dual lattice, the partition function is a single product of  two such holomorphic functions, not a sum of such terms. This is applied to the heterotic string to give a world-sheet action whose quantisation is modular invariant and free from anomalies.
  We discuss modular invariance  for the moduli of both metrics and the extension to higher genus Riemann surfaces.}
 
\author[a]{Chris Hull}

\author[b]{and Neil Lambert}

\affiliation[a]{Blackett Laboratory, Department of Physics, Imperial College, London, UK}
\affiliation[b]{Department of Mathematics, King's College London,\\
 London, UK}

\emailAdd{c.hull@imperial.ac.uk}
\emailAdd{neil.lambert@kcl.ac.uk}

\preprint{Imperial-TP-2025-CH-5}

\begin{document}
 
\maketitle

\section{Introduction and Summary}

In  \cite{Sen:2015nph,Sen:2019qit} Sen gave an interesting action for $2k$-form gauge fields in $4k+2$ dimensions with self-dual field strength.
The degrees of freedom of Sen's theory consist of the physical  $2k$-form gauge field $A$ with self-dual field strength $F$ that couples to the spacetime metric $g$ and the other physical fields, together with a second 
$2k$-form gauge field $C$ with field strength $G$ that doesn't couple to the spacetime metric $g$ and the other physical fields. Instead, $C$ couples to a flat metric $\eta$ and $G$ is self-dual with respect to  $\eta$. Thus the theory consists of the physical chiral gauge field $A$ plus a second chiral gauge field $C$ which decouples from the physical theory. The theory has a diffeomorphism-like symmetry that acts on $F$ but not $G$.
The interaction with the space-time metric $g$ is governed by a tensor $M$ which was constructed perturbatively in  \cite{Sen:2015nph,Sen:2019qit}.
This theory has been further explored in \cite{Lambert:2019diy,Andriolo:2020ykk,Barbagallo:2022kbt,Vanichchapongjaroen:2024tkj,Aggarwal:2025fiq,Hull:2025yww}.  Other   approaches to chiral gauge fields are discussed in e.g.\ \cite{Siegel:1983es,Floreanini:1987as,Devecchi:1996cp,Pasti:1996vs,Belov:2006jd,Mkrtchyan:2019opf,Arvanitakis:2022bnr} and references therein.

In \cite{Hull:2023dgp} a generalisation of Sen's action was constructed in which the non-physical gauge-field $C$ couples to an arbitrary second metric 
$\bar g$ instead of the flat metric $\eta$, and $G$ is self-dual with respect to  $\bar g$.
The theory then describes a physical sector consisting of the gauge field $A$, the spacetime metric $g$ and the other physical fields, together with a shadow sector consisting of the gauge-field $C$ and the second metric $\bar g$. The shadow sector decouples from the physical sector.
This has a number of advantages over Sen's formulation. First, it can be used on any spacetime manifold, not just those admitting a flat metric $\eta$.
Second, it allows a geometric construction of the tensor $M$ in closed form (using results from \cite{Andriolo:2020ykk}). Indeed, finding the tensor $M$, which has a complicated dependence on both metrics  $ g$ and $\bar g$ was a  non-trivial part of the the construction of  \cite{Hull:2023dgp}. Next, it has two gauge symmetries with vector parameters, one for which $ g$ is the gauge field  and one for which $\bar g$ is the gauge field. The diffeomorphism symmetry is a diagonal subgroup of these two symmetries. A related use of a bi-metric theory is proposed for string field theory in \cite{Hull:2025mtb}.

 Both Sen's formulation and its bi-metric generalisation are quadratic in the dynamical fields and so are straightforward to quantise. The purpose of this paper is to explore the quantisation of a chiral scalar in 2 dimensions (the case $k=0$) using the action of \cite{Hull:2023dgp}, 
 with $A$ and $C$ chiral scalars.
 In particular, we   calculate the partition function and show that it has the desired form, which is a holomorphic function of the modulus $\tau$ if the scalar is periodic. Similar calculations were performed before in the case of a torus where one can use the original Sen action with $\bar g =\eta$ \cite{Andriolo:2021gen,Lambert:2023qgs}. 

The physical metric can be written in the form
 \begin{align}\label{g}
   g  = e^{2\omega}(dx+\tau dy)(dx+
 \tilde \tau dy)	 .
 \end{align}
 where $\tau$ and $ \tilde \tau$ are independent real functions for Lorentzian signature and are  conjugate complex functions for Euclidean signature.
 The action of \cite{Hull:2023dgp} describes chiral bosons in Lorentzian signature. However, for the quantum theory we are interested in the path integral for Euclidean signature metrics. In particular, for the partition function we are interested in the case in which the 2-dimensional space is a 2-torus with metric $g$ (\ref{g}) for which $\tau$ is a complex constant, the modulus, with complex conjugate $\tilde \tau$.

The shadow metric $\bar g$ can be written in the same form
 \begin{align}\label{barg}
 \bar g  = e^{2\sigma}(dx+\rho dy)(dx+
 \tilde \rho dy)	 .
 \end{align}
where as before $\rho$ and $ \tilde \rho$ are independent real functions for Lorentzian signature and are  conjugate complex functions for Euclidean signature. We will proceed by using the action of \cite{Hull:2023dgp}  with {\it complex metrics} $ g$ and $\bar g$, so that the real  Lorentzian and Euclidean metrics arise as special cases. Specifically, we will take 
the metrics $ g$ and $\bar g$ to be given by (\ref{g}) and (\ref{barg}), but will consider $\tau$ and $ \tilde \tau$ and
$\rho$ and $ \tilde \rho$ to be {\it independent} complex variables. 
For a 2-torus, the symmetries can be used to set $\tau, \tilde \tau$ and
$\rho, \tilde \rho$ to complex constants, while for higher genus they are functions on the Riemann surface that depend on the moduli.
We will calculate in regions of the  space of complex variables
$\tau, \tilde \tau,\rho, \tilde \rho$ for which the functional integral is well-defined and then analytically continue the results to the whole parameter space.
Complex metrics have also recently been considered in the context of quantum calculations in e.g.\  \cite{Kontsevich:2021dmb,Witten:2021nzp,BenettiGenolini:2025jwe}.
The scalar fields remain real in our calculations.

In two dimensions, the M-tensor that incorporates all dependence on the metric $g$ has one independent component, which we will denote by $M_{--}$ in what follows. For a given metric $g$, we will choose $\bar g$ to be such that $M_{--}$ is finite and non-zero 
everywhere; this is helpful in  calculating the path integrals. We will then analytically continue the result for general $g,\bar g$.

We find a strikingly elegant and simple result. 
For a theory of $8k$ chiral bosons associated with  a torus $\mathbb{R}^{8k} /\Delta$ where $\Delta$ is an even self-dual $8k$-dimensional lattice,
the partition  function
depends on just two of the four moduli and has the factorised form
\begin{equation}\label{Z}
Z(\tau,\rho)= F(\tau) \tilde F (\rho)  . 
\end{equation} 
This has immediate application to   the heterotic string, which involves 16 chiral bosons and an even self-dual lattice which is either the 
root lattice of $E_8
\times E_8$ or the weight lattice of $Spin (32)/\mathbb{Z}_2$.
Then the bi-metric chiral boson theory can be used for the heterotic string where further world-sheet fields coupling to $g$ cancel the modular anomaly
coming from $F(\tau)$, so that the full theory is invariant under $\tau$ modular transformations.
We will see how invariance under $\rho$ modular transformations could also  be achieved by introducing further world-sheet fields coupling to $\bar g$. 

More generally, for a  chiral boson on a circle or for $n$ chiral bosons on a torus, the partition function again depends on just two of the four moduli, 
$\tau$ and $\rho$. In general it is of the form
\begin{equation}\label{Z}
Z(\tau,\rho)= F_o(\tau) \tilde F_o (\rho) f(\tau,\rho) .
\end{equation}
The contribution from the oscillators gives a factorised form $F_o(\tau) \tilde F_o (\rho) $ while the integration over the momenta and winding gives a function $f(\tau,\rho)$ which in general doesn't factorise. 
For  the special case in which the compactification defines a rational conformal field theory (e.g.\ for a boson on a circle of radius $R$ with $R^2$ rational)
then $f(\tau,\rho)$ is a finite sum of terms each of which factorise, so that the partition function takes the form
\begin{equation}
Z(\tau,\rho)\sim
\sum _{a,\bar b} C_{a\bar b}\chi_a(\tau) \chi_{\bar b}(\rho) ,
\end{equation} 
for some constants $C_{a\bar b}$ and some functions $\chi_a(\tau),\chi_{\bar b}(\rho)$ (which are Verma module characters).

This arises in an interesting way. A central role in our analysis is played by an effective action for a non-chiral  scalar coupled to a complex metric
\begin{align}
h_{\mu\nu}dx^\mu dx^\nu  
&=  e^{2\sigma'} (dx+  \rho dy)\left( dx+\tau dy\right) ,
\end{align}
with moduli $\rho,\tau$.
We quantise the scalar coupled to this complex metric by doing the partition function in regions of the $\rho,\tau$ moduli space in which the functional integral is well-defined and then analytically continuing.

Chiral bosons arise in
 a number of different situations.
On 2-dimensional Minkowski space we can have chiral bosons which are non-compact (taking values on a line) or compact (taking values on a circle). For fields on a 2-dimensional cylinder, as arise in closed string theory, a chiral boson with non-trivial momentum can only be achieved for compact bosons, and moreover the radius $R$ of the target space circle must satisfy $R^2=1$ (in units in which the string tension is $1/2\pi$, i.e. $\alpha'=1$). For multiple chiral bosons, they must take values in a rational torus. This will be discussed in detail later.
 However, 
the bi-metric chiral boson action is defined for any 2-dimensional space and any flat target space, either Euclidean space, or a torus which does not need to be rational, and the resulting partition function can have complicated dependence on both $\tau $ and $\rho$. However, for the case of a rational torus 
the theory simplifies and the partition function is a sum of a finite number of terms, each of which factorises into a function of $\tau $ times a function of $\rho$.

One of our main motivations is in the application to string theory where one would want to study chiral bosons on an arbitrary Riemann surface. The bi-metric action is well-suited for this and we study both the torus case and higher-genus Riemann surfaces here. We will discuss the applications to heterotic string theory correlators elsewhere.

The rest of this paper is organised as follows. In section 2 we discuss the action in detail and evaluate it in the special case in which the metrics $g$ and $\bar g$ are given by (\ref{g}) and (\ref{barg}) respectively. In section 3 we turn to the quantum theory. We first discuss the canonical quantization, generalised to a complex metric, and obtain the partition function. We then compute in detail the path integral over a torus to find the same expression.  In section 4 we discuss modular invariance and   what  the ranges of the parameters $\rho,\tilde\rho,\tau,\tilde\tau$ must be to ensure convergence of the integrals in the path integral. In section 5 we extend our results to multiple chiral scalars and show how to obtain a partition function that factorises in to a simple product as in (\ref{Z}) and construct a consistent heterotic string theory by including matter which cancel anomalies in both diffeomorphism symmetries. In section 6 we evaluate the path integral on a generic Riemann surface.  Finally, in section 7 we present our conclusions and a discussion of our results.

\section{The Action}

\subsection{The Action For Chiral Gauge Fields}

The Sen action \cite{Sen:2015nph,Sen:2019qit} is defined in $4k+2$ dimensions. It's generalisation in the presence of a source $J$ and coupled to a general background metric is  \cite{Hull:2023dgp}  
\begin{align}\label{action}
S = - \int \Big(&\frac12 dP\wedge \bar\star  dP - 2 Q\wedge dP-(Q+J)\wedge {\cal M}(Q+J)\nonumber\\
&+2Q\wedge J - \frac12 J\wedge \bar\star J\Big)	 .
\end{align}
Here $P$ is a $2k$-form and $Q$ is  a $(2k+1)$-form which is self-dual with respect to a Lorentzian-signature metric $\bar g$; $Q=\bar\star Q$.  Furthermore, ${\cal M}$ is linear map constructed to take $\bar \star$-self-dual forms to $\bar \star$-anti-self-dual forms and in particular satisfies:
\begin{align}
{\cal M}(Q) = \frac12(1-\bar\star)	{\cal M}\left(\frac12(1+\bar\star)Q\right) .
\end{align}
It is also taken to be symmetric in the sense that
\begin{align}
Q_1\wedge {\cal M}(Q_2) = 	Q_2\wedge {\cal M}(Q_1),
\end{align}
which in particular implies that
$\delta (Q\wedge {\cal M}(Q)) = 2\delta Q\wedge {\cal M}(Q)$, which is useful in deriving the   equations of motion.
For the general construction of ${\cal M}$ see \cite{Sen:2019qit,Andriolo:2020ykk,Hull:2023dgp}.

The role of $ {\cal M}(Q) $ is to ensure that
\begin{align} 
Q + {\cal M}(Q) = \star \left(Q + {\cal M}(Q)\right)	 ,
\end{align}
where $\star $ is the Hodge dual associated to a second, physical  metric $g$ that is  independent of $\bar g$. 
The equations of motion of the action (\ref{action}) take the form
\begin{align}
0 &=   d\left(\bar\star dP+2Q\right) \nonumber\\
0 &= -dP+\bar \star dP - 	2{\cal M}(Q+J) +(1-\bar \star )J  .
\end{align}
We define
\begin{align}
F &= Q+\tfrac12(1+\bar\star )J+{\cal M}(Q+J) \nonumber\\ G &= Q+  \frac12(1+\bar\star )dP  , 
\end{align}
so that $F$ is self-dual with respect to $g$ while $G$ is self-dual with respect to $\bar g$:
\begin{equation}
F =  \star F, \qquad G=\bar \star G .
\end{equation}
The equations of motion can be rewritten as
\begin{align}
dG & = 0\nonumber\\
dF & =  dJ  .
\end{align}
In this way, we find two dynamical form fields,
one self-dual  with respect to the one of the metrics and the other self-dual with respect to the other metric.
  However, only $F$ couples to the source and to the physical metric $g$. While the source $J$ is kept arbitrary here,
   in other examples it takes on an explicit form in terms of various matter fields, for example in terms of other form fields and fermions in  type IIB supergravity. 

\subsection{The action For Chiral Scalars}

In this paper we will focus on the simplest example of two-dimensions, where $P$ is a scalar and $Q$ a 1-form.
It is helpful to use a zweibein frame where (see the appendix for our conventions and notation)
\begin{align}
	{\bar g}_{\mu\nu} &= {\bar e}_\mu{}^a{\bar e}_\nu{}^b\eta_{ab} \nonumber\\ &= {\bar e}_\mu{}^+ {\bar e}_\nu{}^- + {\bar e}_\mu{}^-{\bar e}_\nu{}^+ .
\label{act2d}
\end{align} 
In this case  one sees that the self-duality condition on $Q$ simply becomes
\begin{align}
Q = Q_+\bar e^+,	
\end{align}
so the 1-form $Q$ has only one non-vanishing component
 $Q_+(x)$. 
Furthermore, in such a basis we must have that 
\begin{align}
{\cal M}(Q_+ \bar e^+) = M_{--}Q_+ \bar e^- ,
\end{align}
so the symmetric tensor $M_{\mu\nu}$ has only one non-vanishing component,
 $M_{--}(x)$.
 
Rescaling the  action (\ref{action})  by the string tension  we find
 \begin{align}
S &= -\frac{1}{2\pi}\int {\rm det}(\bar e)\Big(  -\partial_+P\partial_-P - 2 Q_+\partial_-P-(Q_++J_+)^2M_{--} \nonumber\\
&\hskip4cm+2Q_+J_- + J_+J_-\Big)	d^2x , \label{SenHact}
\end{align}
where $\partial_\pm  =\bar e_\pm{}^\mu\partial_\mu$.
The field strengths with $J=0$ are
\begin{equation}
G_+=\frac 1 2 \partial _+P+Q_+, \qquad
F _+=Q_+, \qquad F_-=M_{--}Q_+ ,
\end{equation} 
and $G_-=0$, $F=\ast F$.

 As discussed in the introduction, for a given metric $g$ we will restrict our attention to choices of  $\bar g$ so that $M_{--}(x)$ is well-defined and non-zero for all $x$.
We then find that the action (\ref{act2d}) can be rewritten as
 \begin{align}
S 
& = -\frac{1}{2\pi}\int {\rm det}(\bar e)\Big(  -\partial_+P\partial_-P +M_{--}^{-1}\partial_-P\partial_-P
 \nonumber\\
&\hskip3.4cm
-M_{--} (Q_++J_++M_{--}^{-1}(\partial_-P-J_-))^2\nonumber\\
&\hskip3.4cm +2(J_+-M_{--}^{-1}J_-)\partial_-P -J_+J_-+M_{--}^{-1}J_-^2\Big)	d^2x ,
\end{align}
with the notation $\partial_\pm P = \bar e_{\pm}{}^\mu \partial_\mu P$.
To clean up this action let us shift $Q$  and $J$ are define
\begin{align}
 Q'_+ = 	Q_++J_++M_{--}^{-1}(\partial_-P-J_-) , 
\end{align}
and
\begin{align}
{\cal J}_+ = J_+ - M_{--}^{-1}J_-	 .
\end{align}
The action now takes the form
\begin{equation}
S= S_{eff} + \frac{1}{2\pi}\int {\rm det}(\bar e) M_{--} {Q'_+}^2d^2x  ,
\end{equation}
where
\begin{align}\label{Seff}
S_{eff} &= 	\frac{1}{2\pi}\int {\rm det}(\bar e)\Big(   \partial_+P\partial_-P-M_{--}^{-1}\partial_-P\partial_-P   -2{\cal J}_+\partial_-P +{\cal J}_+J_- \Big)	d^2x .
\end{align}
Thus we see that the action splits into two decoupled parts parts: a contribution involving the scalar $P$ alone, coupled to sources, together with a simple quadratic term in the field $ Q'_+$.  In particular the action for $ Q'_+$ involves no derivatives, so that $ Q'_+$ is an auxiliary field and so  can be integrated  out to give a result that is independent of $P$ and the sources.

\subsection{Evaluating $M_{--}$}

Following \cite{Hull:2023dgp}, to construct ${\cal M}$ it is helpful to introduce $G_{ab}$
such that
\begin{align}
g_{\mu\nu} = \bar e_\mu{}^a\bar e_\nu{}^bG_{ab} 	 .
\end{align}
We see that  $ {\det G} =  -{\det g}/\det \bar g$ and this is non-vanishing so we can introduce the inverse $G^{ab}$. This allows us to write
\begin{align}
(\star F)_a	 &= \bar e_a{}^\mu (\star F)_\mu \nonumber \\
& = \sqrt{-{\det G}}\epsilon_{ab}G^{bc}F_c .
\end{align}
From here we find
\begin{align}
\star F_+ = \sqrt{-{\det G}}G^{-+} F_+ +\sqrt{-G}G^{--}F_-\nonumber\\
 \star F_- = -\sqrt{-{\det G}}G^{+-} F_- -\sqrt{-G}G^{++}F_+ .
\end{align}
Thus if $\star F = F$ then
\begin{align}
 F_+ = \sqrt{-{\det G}}G^{-+} F_+ +\sqrt{-{\det G}}G^{--}F_-\nonumber\\
  F_- = -\sqrt{-{\det G}}G^{+-} F_- -\sqrt{-{\det G}}G^{++}F_+ ,
\end{align}
from which we find
\begin{align}
F_-	 = \frac{1-\sqrt{-{\det G}}G^{-+}}{\sqrt{-{\det G}}G^{--}} F_+ ,
\end{align}
and
\begin{align}
F_-	 = -\frac{\sqrt{-{\det G}}G^{++}}{1+\sqrt{-{\det G}}G^{+-}} F_+ .
\end{align}
These two expressions are equal since $G^{-+}=G^{+-}$  and $G^{++}G^{--}-G^{-+}G^{+-}={\det G}^{-1}$. 

In our case we have $F = Q + {\cal M}(Q)$ so that  $F_+ = Q_+$ and $F_- = M_{--}Q_+$. Therefore demanding that $F$ is self-dual with respect to $\star$ we read off that
\begin{align}
	M_{--} = \frac{1-\sqrt{-{\det G}}G^{-+}}{\sqrt{-{\det G}}G^{--}} =- \frac{\sqrt{-{\det G}}G^{++}}{1+\sqrt{-{\det G}}G^{+-}} .
\end{align}
Note that for a two-by-two matrix we have
\begin{align}
G^{ab}  = \frac{1}{{\det G}}\left(\begin{array}{cc}
G_{--} & -G_{-+} \\ -G_{+-}& G_{++} 	
 \end{array}\right) ,
\end{align}
so
\begin{align}
	M_{--} = -\frac{\sqrt{-{\det G}}- G_{+-}}{ G_{++}} = \frac{ G_{--}}{\sqrt{-{\det G}}+G_{-+}} .
\end{align}

To continue, we 
choose a parameterisation of the   two-dimensional metric $\bar g$ and 
give   express    $M_{--}$ in terms of this parameterisation. Without loss of generality, we can parameterise the metric in terms of  three functions $\sigma (x),\rho(x), \tilde \rho(x)$ as\begin{align}
 \bar g  = e^{2\sigma}(dx+\rho dy)(dx+
 \tilde \rho dy)	 .
 \end{align}
A choice of zweibein is
\begin{align}
\bar e^+ = \frac{1}{\sqrt{2}}e^{\sigma}(dx+\rho dy) ,\qquad  	\bar e^- = \frac{1}{\sqrt{2}}e^{\sigma}(dx+\tilde \rho dy) .
\end{align}
Then $\det \bar g =-(\det \bar e)^2$ with
\begin{align}
\det \bar e  =  \frac12 e^{2\sigma}(\tilde\rho-\rho) 	 .
\end{align}
 Similarly for the physical metric we write
 \begin{align}
   g  = e^{2\omega}(dx+\tau dy)(dx+
 \tilde \tau dy)	 ,
 \end{align}
 for other functions $\omega(x),\tau(x),\tilde \tau(x)$.
 In this case we take
 \begin{align}
\bar e^+ = \frac{1}{\sqrt{2}}e^{\omega}(dx+\tau dy) ,\qquad  	\bar e^- = \frac{1}{\sqrt{2}}e^{\omega}(dx+\tilde \tau dy) ,
\end{align}
and find $\det   g =-(\det  e)^2$ with
\begin{align}
  \det e  =  \frac12 e^{2\omega}(\tilde\tau- \tau) 	 .
\end{align} 
 A short computation shows that
 \begin{align}
 G_{ab} = 2\frac{e^{2(\omega-\sigma)}}{(\rho-\tilde \rho)^2}	\left(\begin{array}{cc}
(\tilde \rho-\tau)(\tilde \rho-\tilde\tau) &-\tau\tilde\tau -\rho\tilde\rho+\tfrac12(\tau+\tilde \tau)(\rho+\tilde\rho) \\-\tau\tilde\tau -\rho\tilde\rho+\tfrac12(\tau+\tilde \tau)(\rho+\tilde\rho) & (  \rho-\tau)(  \rho-\tilde\tau)\\
\end{array}
\right) ,
 \end{align}
and hence we obtain
 \begin{align}
M_{--} 
&= 	\frac{(\rho-\tau)(\rho-\tilde\tau)}{\tfrac12(\rho-\tilde\rho)(\tau-\tilde\tau )-\tau\tilde\tau -\rho\tilde\rho+\tfrac12(\tau+\tilde \tau)(\rho+\tilde\rho) }\nonumber\\
&=-
\frac{ \rho-\tau  }{  \tilde\rho-\tau } .
\end{align}
In the special case $\rho=-\tilde\rho =1$ we have
\begin{align}
M_{--} 
& =  \frac{(1-\tau) }{ (1+\tau) } .
\end{align}
This  agrees with a similar computation in  \cite{Andriolo:2021gen} (up to $\tau\to-1/\tau$  and a change in sign arising from a redefinition of ${\cal M}$).

Let us make some comments about $M_{--}$ in special cases. First we note that when $\rho=\tau$ we find $M_{--}=0$. This arises as the one form $dx+\rho dy=dx+\tau dy$ is then self-dual with respect to both $g$ and $\bar g$. If we also impose $\tilde \rho=\tilde \tau$,
 then $g$ is conformal to $\bar g$. Since the self-duality condition is conformally invariant we expect to find $M_{--}=0$ in this case. However, we also find $g$ is conformal to $\bar g$ if we take  $\tilde \rho=\tau$ and $\tilde \tau=\rho$ but this corresponds to $M_{--}\to \infty$. The reason for this
 is
  that while $dx+\rho dy$ is always taken to be self-dual with respect to $\bar g $, $dx+\tilde\tau dy = dx + \rho dy$ is anti-self-dual with respect to $g$. Thus  although the two metrics are conformal to each other the notion of self-duality has been flipped.
  As discussed in the introduction, we choose $\bar g$ so that $M_{--}$ is finite and non-zero everywhere, so for given   $\tau, \tilde \tau$ we choose
  $\rho, \tilde \rho$ to exclude the above cases.
  
 Note that the classical theory has two Weyl symmetries, one for each metric,  under
 \begin{equation}
g _{\mu\nu} \to \Omega^2 (x) g _{\mu\nu}
,
\qquad
\bar g _{\mu\nu} \to \bar\Omega^2 (x)\bar g _{\mu\nu}
\end{equation}
as can be seen from the fact that $M$ doesn't depend on $\omega$ or $\sigma$.

\subsection{Evaluating $S_{eff}$}

We now discuss further the effective action (\ref{Seff}) that we found for $P$. It is helpful to write it in a more familiar form as
\begin{align} 
S_{eff} &= 	\frac{1}{2\pi}\int \sqrt{-\det h}\Big( \frac12h^{\mu\nu}  \partial_\mu P \partial_\nu P -2{\cal J}^\mu \partial_\mu P +{\cal J}^\mu J_\mu \Big)	d^2x ,
\end{align}
where
\begin{align}
	 \sqrt{-\det h}  h^{\mu\nu}   
	 &= \det\bar e( \bar e_+{}^{\mu}\bar e_-{}^{\nu}+\bar e_-{}^{\mu}\bar e_+{}^{\nu}-2M_{--}^{-1}\bar e_-{}^\mu \bar e_-{}^\nu)   ,
\end{align}
and
\begin{align}
{\cal J}^\mu = \frac{\det \bar e}{\sqrt{-\det h}}\bar e_-{}^\mu{\cal J}_+ .
\end{align}
Note  $h_{\mu\nu}$ is only determined  up to a conformal factor that drops out of the action due to conformal invariance. A possible zweibein frame is
\begin{align}
	h_{\mu\nu} = \tilde e_\mu{}^a \tilde e_\nu{}^b\eta_{ab} ,
\end{align}
with
\begin{align}
\tilde e_\mu{}^+ &= 	\bar e_\mu{}^{+}\qquad \tilde e_\mu{}^- = \bar e_\mu{}^- +M_{--}^{-1} \bar e_\mu{}^+\\
\tilde e_+{}^\mu &= 	\bar e_+{}^\mu -M_{--}^{-1} \bar e_-{}^\mu
\qquad \tilde e_-{}^\mu = \bar e_-{}^\mu . \end{align}
Its easy to see that  $\det \tilde e=\det \bar e$ and hence $\sqrt{-\det h}=\det \bar e$ and  ${\cal J}^\mu =  \bar e_-{}^\mu{\cal J}_+ = \tilde e_-{}^\mu {\cal J}_+$. 
In particular. given the metric parameterizations above we find
\begin{align}
	 \tilde e^+ & = \frac{1}{\sqrt2}e^\sigma(dx+  \rho dy)\\
	 \tilde e^-& = \frac{1}{\sqrt2}e^\sigma\frac{\rho-\tilde\rho}{\rho-\tau}\left( dx+\tau dy\right) .
\end{align}
By a local Lorentz transformation 
\begin{align}
\tilde e'^+ = \Lambda \tilde e^+	\qquad \tilde e'^- = \Lambda^{-1}\tilde e^- ,
\end{align}
with $\Lambda= \sqrt{(\rho-\tau)/(\rho-\tilde\rho)}$, 
we can also   find something more symmetric:
\begin{align}
	 \tilde e'^+ & = \frac{1}{\sqrt2} e^{\sigma'} (dx+  \rho dy)\\
	 \tilde e'^-& = \frac{1}{\sqrt2} e^{\sigma'} \left( dx+\tau dy\right) .
\end{align}
For either choice of the zweibein the  metric $h$ can be taken to be
\begin{align}
h_{\mu\nu}dx^\mu dx^\nu  
&=  e^{2\sigma'} (dx+  \rho dy)\left( dx+\tau dy\right) ,
\end{align}
where we have introduced $ \sigma' =\sigma + \tfrac12\ln((\rho-\tilde\rho)/(\rho-\tau))$. However, we recall that the metric $h$ is only defined up to a conformal factor so we can choose $\sigma'$ to be anything we like.

\section{The Quantum Partition Function}

\subsection{The Path Integral And A Complex Metric}

Let us now turn our attention to the quantum theory. We start with the path integral 
\begin{align}\label{Zdef}
Z[J] & = \int [dP][d Q_+] e^{iS[P,Q_+,J]}\nonumber\\
  &= \int [dP][d Q'_+] e^{iS_{eff}[P,J] + \frac{i}{2\pi} \int {\det }(\bar e)M_{--}(Q'_+)^2} .
\end{align}
Our first task is to perform the path integral over $ Q'_+$, which is a  simple Gaussian  
giving formally an infinite factor  $\prod _x ({\det }(\bar e)M_{--})^{-1/2}(x)$. In \cite{Andriolo:2021gen} it was argued that
this can be regularised to give a finite number that does not depend on   ${\det }(\bar e)M_{--}$. 
Dropping this  irrelevant  multiplicative factor, we    find
\begin{align}
Z[J]
   \sim \int [dP] e^{iS_{eff}[P,J]} .	
\end{align}

In \cite{Andriolo:2021gen}   the path integral was made convergent   by taking the metric $g$ to   have Euclidean signature. 
Here, we wish to consider the path integral for the case in which both metrics $\bar g$ and $g$ have Euclidean signature.
As discussed in the introduction, we actually consider the more general case in which $\bar g$ and $g$ are complex metrics, with
$\rho,\tilde\rho, \tau$ and $\tilde \tau$ independent  complex variables.
We then find Riemannian metrics  $\bar g$ and $g$ in the case that $\tilde\rho = \rho^*$ and $\tilde\tau = \tau^*$ respectively, but we can also be more general. 
 In what follows, we assume that spacetime is a compact Riemann surface; additional subtleties  can arise on non-compact spaces.

To continue, we write  $\sqrt{-h}=i\sqrt{h}$  and $S_{eff}=i\tilde S_{eff}$ with
 \begin{align}
  \tilde S_{eff}=	\frac{1}{2\pi}\int \sqrt{h}\Big( \frac12h^{\mu\nu}  \partial_\mu P \partial_\nu P -2{\cal J}^\mu \partial_\mu P +{\cal J}^\mu J_\mu \Big)	d^2x .	
 \end{align} 
so that now  the path integral takes the form
\begin{align} 
  Z[\rho, \tau ,J]  &\sim  \int [dP]e^{-  \tilde  S_{eff}[P,J]} , \end{align}
which depends holomorphically on each of the two complex structures, $\rho$ and $\tau$. 
 This is the same as a path integral for a non-chiral real scalar field but with a complex metric:
 \begin{align}
 h = e^{2\sigma'}(dx+\rho dy)(dx+\tau dy)	 .
 \end{align}
Here we see a complex metric combining the holomorphic sectors of two distinct Riemannian metrics
arising naturally from the Sen action.

\subsection{A Canonical Approach}

It is instructive to construct the partition function from  canonical quantization on a 2-torus. As usual, for real Riemannian metrics, 
the diffeomorphism symmetry can be used to bring the real metric $g$ on a 2-torus to standard form with $\tau$ constant, the complex structure modulus. 
 As explained in \cite{Hull:2023dgp}, there is a second diffeomorphism-like symmetry that can be used to bring  the real metric $\bar g$ to standard form with $\rho$ a constant modulus.  
 Here  we set the sources to zero and   take   the functions $\rho$ and $\tau$ to be  constants. 
As already mentioned we will extend our calculations to allow for  $\rho,\tilde\rho, \tau$ and $\tilde \tau$ to be general independent  complex variables.
 
 The metrics are then (using the symmetries to set the conformal factors to 1)
 \begin{eqnarray}
g&=&(dx + \tau d y)(dx + \tilde\tau d y) ,
\\
\bar g&=&(dx + \rho d y)(dx +\tilde \rho d y) ,
\\
h&=&(dx + \tau d y)(dx + \rho d y) .
\end{eqnarray}
 It is helpful to introduce the coordinates
\begin{align}\label{coords}
z = x + \tau y ,\qquad \tilde z = x+\rho y  .	
\end{align}
so that the metric is simply
\begin{align}
h = dzd\tilde z	 .
\end{align}
 Note that for $\rho=-\tau=1$ we find that $h$ is the Minkowski metric with $y$ the time variable. Here we will be more general and allow for complex values of $\rho$ and $\tau$. The 
associated derivatives are 
\begin{align}\label{derivatives}
 \partial = -\frac{1}{\rho-\tau}\left(\partial_y - \rho \partial_x\right),\qquad \tilde \partial = 	  \frac{1}{\rho-\tau}\left( \partial_y - \tau\partial_x\right) ,
 \end{align}
and  these satisfy
 \begin{align}
 \partial z = \tilde\partial \tilde z =1,\qquad \partial  \tilde z = \tilde\partial z =0 .
 \end{align}
 The equation of motion for $P$ becomes simply
 \begin{align}
 \partial\tilde\partial P =0 ,
 \end{align}
Assuming the coordinate $x$ is periodic: $x\cong x + 2\pi l$, the general solution can be parameterised as
\begin{align}
P = P_0 + \frac1{2l}a_0z - \frac1{2l}\tilde a_0  \tilde z + \frac{i}{\sqrt{2}}\sum_{n\ne0}\frac{a_n}{n}e^{ -inz/l}-\frac{i}{\sqrt{2}}\sum_{n\ne0}\frac{\tilde a_n}{n}e^{ in\tilde z/l}	 .
\end{align}
Note that the linear terms imply that  $P(x+2\pi l,y)=P(x,y)+\pi ( a_0-\tilde a_0)$. 

There are two cases to consider. First, if $P$ is a non-compact boson $P\in\mathbb R$, then we must have $a_0=\tilde a_0$ so setting
\begin{align}
a_0 = p_0 ,\qquad \tilde a_0 = p_0  
\end{align}
we have
\begin{align}
P = P_0 + \frac1{2l}p_0 (\tau-\rho)y + \frac{i}{\sqrt{2}}\sum_{n\ne0}\frac{a_n}{n}e^{ -inz/l}-\frac{i}{\sqrt{2}}\sum_{n\ne0}\frac{\tilde a_n}{n}e^{ in\tilde z/l}	 
\end{align}
with centre of mass position $P_0$ and continuous momentum proportional to $p_0$.

Secondly, if $P$ is periodic with $P\cong P+2\pi R$, then we must have $ a_0-\tilde a_0=2mR$ for some constant $R$ and integer $m\in\mathbb Z$. To this end, we write 
\begin{align}
a_0 = p_0+mR,\qquad \tilde a_0 = p_0-mR	 .
\end{align}
Moreover, the momentum $p_0$ is discrete as the eigenvalues of $p_0$ are quantised: $p_0=n/R$ for integers $n\in\mathbb Z$. Then the states are labelled by two integers $m,n$, with $n/R$ the discrete momenta and $m$ being the winding number.


 To quantize this system we treat $y$ as time and obtain the conjugate momentum
 \begin{align}
 \Pi &= \frac{i}{2\pi}\sqrt{h}h^{y\nu}\partial_\nu P \nonumber\\ 
 &= 	 \frac{1}{2\pi}  (\partial-\tilde\partial)P\nonumber\\
 & =  \frac{1}{2\pi l} \left(p_0  + \frac{1}{\sqrt{2}}\sum_{n\ne0}a_ne^{ -inz/l}-\frac{1}{\sqrt{2}} \sum_{n\ne0} {\tilde a_n} e^{ in\tilde z/l}	\right)  .\end{align}
To reproduce the quantization condition
\begin{align}
[P(x,y),\Pi(x',y)] = i\delta(x-x')	 ,
\end{align}
we find the non-zero commutation relations
\begin{align}
[a_n,a_m] = n\delta_{m,-n},\qquad 	[\tilde a_n,\tilde a_m] = n\delta_{m,-n} ,\qquad [P_0,p_0]=i .
\end{align}

The Hamiltonian is
\begin{align}
H &=  	\int \Pi \partial_yP - {\cal L}dx \nonumber\\
& = \frac{1}{2\pi}\int\left(\rho (\tilde\partial P)^2-\tau(\partial P)^2 \right)dx \nonumber\\
 & = \frac{\rho}{l} H_\rho-\frac{\tau}{l} H_\tau ,
\end{align}
where
\begin{align}
H_\tau &=  \frac14(p_0+mR)^2 +\sum_{n>0} a_{-n}a_{n}	-\frac{1}{24} \nonumber\\
H_\rho &=  \frac14(p_0-mR)^2+\sum_{n>0} \tilde a_{-n}\tilde a_{n}-\frac{1}{24}  ,
\end{align}
and as usual the $-1/24$ arises from  normal ordering.

As is familiar in two dimensions, the left and right sectors decouple so that the total Hilbert space is a direct product
 ${\cal H} = {\cal H}_\tau\otimes {\cal H}_\rho$.
   Here this corresponds to the decoupling of the two chiral bosons, as first observed in \cite{Sen:2019qit}. 
Since the Hilbert space factorises   the partition function takes the form
\begin{align}
Z(\beta) &= {\rm Tr}_{{\cal H} }\left( e^{-   \beta  H} \right)\nonumber\\ &={\rm Tr}_{{\cal H}_\tau\otimes {\cal H}_\rho}\left( e^{ \beta l^{-1} \tau H_\tau}e^{-\beta l^{-1}\rho H_\rho}\right)\nonumber\\
   &={\rm Tr}_{{\cal H}_\tau\otimes {\cal H}_\rho} \left(q^{H_\tau} \tilde q^{H_\rho}\right) ,
\end{align}
where $q = e^{\beta l^{-1}\tau}$ and $\tilde q = e^{-\beta l^{-1} \rho}$.
This can be evaluated in the usual way. Summing over the oscillators gives
\begin{align}
Z(\beta) = 	\frac{1}{\eta(q)\eta(\tilde q)}{\rm Tr}_{zero-modes} \left(q^{H_\tau} \tilde q^{H_\rho}\right) 
\end{align}
and it remains to take the trace of $q^{H_\tau} \tilde q^{H_\rho}$ over the zero-modes.
 If $P$ is a non-compact boson,    then  there are no winding numbers ($m=0$) and the momentum $p_0$ is a continuous variable that we integrate over to find
\begin{align}
Z(\beta) \sim	\frac{1}{\sqrt{\rho-\tau}}\frac{1}{\eta(q)\eta(\tilde q)}  .\end{align}
On the other hand, if $P$ takes values in a circle of radius $S$,
  then the eigenvalues of $p_0$ are quantised: $p_0=n/R$. Summing over $n$ and $m$ leads to 
\begin{align} \label{partfun}
Z(\beta) \sim 	 \frac{1}{\eta(q)\eta(\tilde q)}\sum_{m,n} q^{\frac14(n/R+mR)^2} \tilde q^{\frac14(n/R-mR)^2}  .\end{align}
We will find agreement with the path integral calculation below by taking $\beta = 2\pi i l$, as expected.

\subsection{Path Integral On The Torus}

Our next task is to provide an explicit calculation of the partition function by evaluating the path integral on a torus.  As above we take 
\begin{align}
h =  (dx+\rho dy)(dx+\tau dy)	 ,
\end{align}
with $\rho$ and $\tau$ constant but complex and unrelated. We assume that $x,y\in [0,2\pi l]$ with periodic boundary conditions (except for the zero-modes). In this section we will closely follow the calculations in sections 10.2 and 10.4 of \cite{DiFrancesco:1997nk}. The set-up discussed in \cite{DiFrancesco:1997nk} arises  in our analysis on setting  
 $\rho = \tau^*$, but here we will keep $\rho$ and $\tau $ unrelated. A related calculation for the Sen action with $\rho=1$ was performed in \cite{Andriolo:2021gen} and also \cite{Lambert:2023qgs}.

 First we split $P=P_{z.m.}+ P_{o}$ where $P_{o}$ represents the non-zero oscillator modes which can be expanded in a basis of eigenmodes
 \begin{align}
 	P_{o}=\sum_{n,m} c_{n,m}\phi_{n,m} ,
 \end{align}
where 
 \begin{align}
 -\frac{1}{\sqrt{h}}\partial_\mu\left(\sqrt{h}h^{\mu\nu}\partial_\nu\phi_{n,m}\right) = \lambda_{n,m}\phi_{n,m}	 ,
 \end{align}
 where
 $\sqrt{h}=\sqrt{det(h_{\mu\nu})}$
and we normalise
\begin{align}
\int d^2x \sqrt{h} \phi_{m,n}\phi_{-m',-n'}	 =	\delta_{n,n'}\delta_{m,m'} .
\end{align}
Taking a Fourier expansion we find
\begin{align}
\phi_{m,n} = N e^{i m x/l+iny/l}	 ,
\end{align}
with $N^{-2}=4\pi^2 l^2\sqrt{h}  $ a normalization constant and   \begin{align}
\lambda_{m,n} =  - \frac4{l^2}\frac{1}{(\rho-\tau)^2}(n-m\rho)(n-m\tau)	 .
\end{align}
The reality of $P$ is imposed by taking $c^*_{m,n}=c_{-m,-n}$. 

The path integral leads to an infinite number of Gaussian integrals
\begin{align}
Z_{o}(\tau,\rho) & = \prod_{m,n}\int dc_{m,n}  e^{-\frac{1}{4\pi}|c_{m,n}| ^2\lambda_{m,n} + ic_{m,n}j_{-m,-n}	} ,
\end{align}
 over the complex coefficients $c_{n,m}$ with $(m,n)\ne (0,0)$. Here $j_{m,n}$ arises from the sources:
 \begin{align}
 { J}_\mu &= \sum_{m,n
 } { J}_{\mu m,n} e^{imx/l+iny/l}\nonumber\\
 	j_{m,n} &= 2\pi l\frac{\rho m-n}{\rho-\tau}(\tau { J}_{xm,n} - { J}_{ym,n}) .
 \end{align}
  The reality condition $c^*_{n,m}=c_{-n,-m}$ allows us to write this as
 \begin{align}
Z_{o}(\tau,\rho)  & = \prod_{(m,n) >0}\int dc_{m,n}dc^*_{m,n}  e^{-\frac{1}{4\pi}c_{n,m}c^*_{n,m}\lambda_{m,n} + ic_{m,n}j_{-m,-n}	+ ic^*_{m,n}j_{m,n}}	\nonumber\\
& = \prod_{(m,n) >0}\int dc'_{m,n}d{c'}^*_{m,n}  e^{-\frac{1}{4\pi}c'_{n,m}{c'}^*_{n,m}\lambda_{n,m}+4\pi \lambda_{m,n}^{-1}j_{m,n}j_{-m,-n}} ,
\end{align}
where we have restricted the product to be only  over positive  modes 
(for example  only those pairs $(m,n)$ whose first non-zero entry is positive) and in the second line we have shifted the integration variables to remove the linear term.
Thus we find 
\begin{align}
Z_{o} = e^{-S_{n.l.}}\prod_{(m,n)>0}\frac{4\pi^2}{\lambda_{m,n}}	  , 
\end{align}
where 
\begin{align}
	e^{-S_{n.l.}} = e^{4\pi\sum_{(m,n) >0} \lambda_{m,n}^{-1}j_{m,n}j_{-m,-n}} .
\end{align}
However, since $\lambda_{n,m}$ is even in $(n,m)\to (-n,-m)$ we will find it easier to write this as
\begin{align}
Z_{o}(\tau,\rho) = e^{-S_{n.l.}}{\prod_{m,n}}'\frac{2\pi}{\sqrt{\lambda_{m,n}}}	 ,
\end{align}
where the product is over all $n,m\in \mathbb Z$, not both zero. 
We use the notation ${\prod_{m,n}}'$ or ${\sum_{m,n}}'$
to indicate a product or summation over all integers $m,n$ where not both are zero.
As usual, to regulate this expression we  define
\begin{align}
G(s) &= (4\pi^2)^s{\sum_{m,n}}'\lambda_{m,n}^{-s}\\
& =  {(4\pi^2l^2\sqrt{h})^{s}}	{\sum}'_{m,n}\frac{1}{(n-m\rho)^s(n-m\tau)^s	} ,
\end{align}
and identify
\begin{align}
Z_{o}(\tau,\rho) =e^{-S_{n.l.}} e^{\frac12 G'(0)}	 .
\end{align}
Next, we write
 \begin{align}
G(s)  
& =  {(4\pi^2l^2\sqrt{h})^{s}}	\tilde G(s) ,
\end{align}
so that we  then have  
\begin{align}\label{preF}
Z_{o} =   (4\pi^2l^2\sqrt{h} )^{\frac12\tilde G(0)} e^{\frac12 \tilde G'(0)}	e^{-S_{n.l.}} .
\end{align}
In other words, the coefficient  $4\pi^2l^2\sqrt{h}$ simply provides an overall  factor that multiplies the regulated infinite product. 

Our next task is to compute $\tilde G(s)$, which exists for suitably large $s$, and with certain restrictions on $\rho$ and $\tau$, and then analytically continue it to $s=0$. 
In particular we find
\begin{align}
\tilde G(s) & ={\sum_{ n,m }}'	\frac{1}{(n-m\rho)^s(n-m\tau )^s	} \nonumber \\
& = \sum_{n\ne 0}\frac{1}{ n^{2s}} + \sum_{m\ne 0 }\sum_n	\frac{1}{(n-m\rho)^s(n-m\tau )^s	}\nonumber\\
& = (1+(-1)^{2s})\zeta(2s) +  \sum_{m\ne 0 }\sum_n	\frac{1}{(n-m\rho)^s(n-m\tau )^s	} .
\end{align}
To continue, we write 
\begin{align}
\chi = \frac12(\rho +\tau )	, \qquad \varphi = \frac12(\rho-\tau) ,
\end{align}
so that 
\begin{align}
\tilde G(s)   = (1+(-1)^{2s})\zeta(2s) +  \sum_{m\ne 0 }\sum_n	\frac{1}{(n-m\chi-m\varphi)^s(n-m\chi+m\varphi )^s	} .
\end{align}
Here we notice that for fixed $m$  the sum  over $n$ is periodic under $m\chi\to m\chi+1$ so we can make a Fourier expansion
 \begin{align}
\tilde G(s)   = (1+(-1)^{2s})\zeta(2s) +  \sum_{m\ne 0 }\sum_n\sum_p e^{2\pi i mp\chi }	\int_0^1 dy\frac{e^{-2\pi i py}}{(n-y-m\varphi)^s(n-y+m\varphi )^s	} .
\end{align}
Next, we note that the sum over $n$ converts the integral over $y$ from $[0,1]$ to $(-\infty,\infty)$:
 \begin{align}
\tilde G(s)   &= (1+(-1)^{2s})\zeta(2s) +  \sum_{m\ne 0 } \sum_p e^{2\pi i mp\chi }	\int_{-\infty}^\infty  dy\frac{e^{-2\pi i py}}{(-y-m\varphi)^s(-y+m\varphi )^s	}\nonumber\\
&= (1+(-1)^{2s})\zeta(2s) +  \sum_{m\ne 0 } \sum_p e^{2\pi i mp\chi }	\int_{-\infty}^\infty  dy\frac{e^{2\pi i py}}{(y-m\varphi)^s(y+m\varphi )^s	}\nonumber\\
&= (1+(-1)^{2s})\zeta(2s) +  \sum_{m\ne 0 } \sum_p e^{2\pi i mp\chi }	\int_{-\infty}^\infty  dy\frac{e^{2\pi i py}}{(y^2-m^2\varphi^2)^s 	} .
\end{align}
We now introduce a new integral:
 \begin{align}
\tilde G(s)  &= (1+(-1)^{2s})\zeta(2s) +  \frac{1}{\Gamma(s)}\sum_{m\ne 0 } \sum_p \int_{-\infty}^\infty  dy\int_0^\infty \frac{dt}{ t^{1-s}}e^{2\pi i p(m\chi+y) }	e^{-t(y^2-m^2\varphi^2)} .
\end{align}
Note that convergence of the $t$ integral requires that ${\rm Re}(\varphi^2) <0$.
The integral over $y$ is Gaussian and can now be computed
\begin{align}
\tilde G(s)  &= (1+(-1)^{2s})\zeta(2s) +  \frac{\sqrt{\pi}}{\Gamma(s)}\sum_{m\ne 0 } \sum_p \int_0^\infty \frac{dt}{ t^{3/2-s}}e^{2\pi i pm\chi  }	e^{ m^2 t\varphi^2-\pi^2p^2/t} .
\end{align}
We need to treat the $p=0$ term separately:
\begin{align}
	\sum_{m\ne 0}\int_0^\infty \frac{dt}{ t^{3/2-s}} e^{ m^2 t\varphi^2 } 
	&= \sum_{m\ne 0} \frac{\sqrt{\pi}}{\Gamma(s)}(-m^2\varphi^2)^{1/2-s}\nonumber\\
	&= 2\frac{\sqrt{\pi}}{\Gamma(s)}(\sqrt{-\varphi^2})^{1-2s}\zeta(s-1/2) ,
\end{align}
which again is valid if ${\rm Re}(\varphi^2) <0$. Now we invoke the $\zeta$-function identity
\begin{align}
\pi^{-s'/2}\Gamma(s'/2)\zeta(s'/2)	= \pi^{(s'-1)/2}\Gamma((1-s')/2)\zeta(1-2s') ,
\end{align}
to find something valid near $s=0$:
\begin{align}
\tilde G(s)  &= (1+(-1)^{2s})\zeta(2s)  + 2\frac{\pi^{2s-1}}{\Gamma(s)}\Gamma(1-s)\zeta(2-2s)(\sqrt{-\varphi^2})^{1-2s}\nonumber \\
&\qquad + \frac{\sqrt{\pi}}{\Gamma(s)}\sum_{m\ne 0 } \sum_{p\ne0}  \int_0^\infty \frac{dt}{ t^{3/2-s}}e^{2\pi i pm\chi  }	e^{ m^2 t\varphi^2-\pi^2p^2/t} .
\end{align}
To continue, we note that for small $s$,  $\Gamma(s)\sim 1/s$ and so we can read off that
\begin{align}
\tilde G(0) = 2\zeta(0)=-1	 .
\end{align}
Furthermore, (noting that $\zeta'(0)=-\tfrac12\ln 2\pi$)
\begin{align}
\tilde G'(0) &= (-\pi i-2\ln 2\pi)+ 	  2 {\pi^{-1}} \Gamma(1)\zeta(2)\sqrt{-\varphi^2}\nonumber \\
&\qquad +  {\sqrt{\pi}} \sum_{m\ne 0 } \sum_{p\ne0} e^{2\pi i pm\chi  }	 \int_0^\infty \frac{dt}{ t^{3/2}}e^{ m^2 t\varphi^2-\pi^2p^2/t} .
\end{align}
This final integral can be computed (and exists for ${\rm Re}(\varphi^2)<0$):
\begin{align}
\tilde G'(0) &= (-\pi i-2\ln 2\pi)+ 	 \frac{\pi }{3}  \sqrt{-\varphi^2}\nonumber \\
&\qquad +   \sum_{m\ne 0 } \sum_{p\ne0}  \frac{1}{|p|} e^{2\pi i pm \chi  -2\pi | p||m |\sqrt{-\varphi^2}} \nonumber\\
& \sim 	 \frac{\pi }{3}  \sqrt{-\varphi^2}+2\sum^\infty_{m=1 } \sum^\infty_{p=1}  \frac{1}{p} e^{2\pi i pm \chi  -2\pi  pm \sqrt{-\varphi^2}}+\sum^\infty_{m=1 } \sum^\infty_{p=1}  \frac{1}{p} e^{-2\pi i pm \chi  -2\pi  pm \sqrt{-\varphi^2}} .
\end{align}
The first term just gives a constant multiplicative factor and we have dropped it in the second line. If we introduce
\begin{align}
	q = e^{2\pi i\chi - 2\pi \sqrt{-\varphi^2}},\qquad \tilde q = e^{-2\pi i\chi - 2\pi \sqrt{-\varphi^2}} ,
\end{align}
then 
\begin{align}
\tilde G'(0)  & \sim 	 \frac{\pi }{3}  \sqrt{-\varphi^2}+2\sum^\infty_{m=1 } \sum^\infty_{p=1}  \frac{1}{p} q^{pm}+\sum^\infty_{m=1 } \sum^\infty_{p=1}  \frac{1}{p}\tilde q^{pm} \nonumber\\
& \sim 	 \frac{\pi }{3}  \sqrt{-\varphi^2}-2\sum^\infty_{m=1 }    \ln(1- q^{m})-2\sum^\infty_{m=1 }  \ln (1- \tilde q^{m}) .\end{align}
Let us recall that $\chi = (\rho+\tau)/2$ and $\varphi = (\rho-\tau)/2$. If we take
\begin{align}
	\sqrt{-\varphi^2} = i(\rho-\tau)/2 ,
\end{align}
then we find
\begin{align}
	q = e^{2\pi i \tau},\qquad \tilde q = e^{-2\pi i \rho} .
\end{align}
Putting all the pieces together, we find
\begin{align}
Z_{o}(\tau,\rho) \sim \frac{1}{\sqrt{\rho-\tau}}\frac{1}{\eta(q)}\frac{1}{\eta(\tilde q)}e^{-S_{n.l.}} ,	
\end{align}
where 
\begin{align}
\eta(q) = q^{1/24}\prod_{m=1}^\infty (1-q^m)	 ,
\end{align}
 and $(\rho-\tau)^{-\frac12}$ comes from the $(4\pi l^2\sqrt{h})^{\frac12\tilde G(0)} = (4\pi l^2\sqrt{h})^{-\frac12}$ prefactor in (\ref{preF}). 
 
Next we look at the zero-mode contribution. Here we allow for $P$ not to be single valued but rather 
to take values on a circle with $P\cong P+2\pi R$ for some fixed $R$  
so that $P$ can have winding modes. Then with winding numbers $m,n$ on the $x$ and $y$ directions we have
\begin{align}
P_{z.m.} = p_0 + \frac{mRx}{l} +  \frac{nRy}{l} 	 .
\end{align}
 Such field configurations satisfy $- h^{\mu\nu}\partial_\mu\partial_\nu P_{z.m.}=0$. Evaluating
the action on them gives
\begin{align}
\tilde S_{z.m.} &= \frac{1}{2\pi}\int  \sqrt{h}\left(\frac12h^{\mu\nu}\partial_\mu P_{z.m.}	\partial_\nu P_{z.m.} - 2 {\cal J}^\mu_{0,0}\partial_\mu P_{z.m.}\right)d^2 x	\nonumber\\
& = \frac{\pi R^2}{\sqrt{h}}(n-\rho m)(n-m\tau) - \frac{4\pi l R}{\rho-\tau} (\tau J_{x0,0}-J_{y0,0})(n-\rho m ) .
\end{align}
Noting that $\sqrt{h}=i(\rho-\tau)/2$,  this leads to 
\begin{align}\label{Zzm}
Z_{z.m.}(\tau,\rho) = \sum_{m,n}e^{-\frac{2\pi i R^2}{ {\rho-\tau}}(n-\rho m)(n-m\tau)-\frac{4\pi   lR}{ {\rho-\tau}}(\tau J_{x0,0}-J_{y0,0})(n-\rho m )
}	 .
\end{align}
Next, we   perform a Poisson re-summation over $n$:
\begin{align}
\sum_n e^{-\pi An^2 + Bn}	= \frac{1}{\sqrt{A}}\sum_k e^{-\frac{\pi}{A}\left(k+\frac{B}{2\pi i}\right)^2} .
\end{align}
This requires that ${\rm Re} A >0$.
Using $A = 2 iR^2/(\rho-\tau)$ and $B=\pi R^2(\rho+\tau)m/\sqrt{h}-4\pi l R(\tau J_{x0,0}-J_{y0,0})/(\rho-\tau)$, we get 
\begin{align}
Z_{z.m.}(\tau,\rho) & =  \frac{\sqrt{\rho-\tau}}{\sqrt{2i}R}\sum_{m,k}e^{\frac{2 \pi i R^2}{{\rho-\tau}}\rho\tau m^2 - \frac{\pi i({\rho-\tau})}{2R^2} k^2 + \pi i (\rho+\tau)mk -\frac{\pi i R^2m^2}{2(\rho-\tau)}(\rho+\tau)^2 }\nonumber \\&\hskip2cm \times e^{2\pi  l \left(ik/R-2i(\rho+\tau)mR/(\rho-\tau)+4im R\rho/(\rho-\tau)\right) (\tau J_{x0,0}-J_{y0,0})  -\frac{2\pi il^2}{\rho-\tau} (\tau J_{x0,0}-J_{y0,0})^2}\nonumber\\
& =  \frac{\sqrt{\rho-\tau}}{\sqrt{2i}R} \sum_{m,k}e^{-\frac{\pi i R^2}{2 }(\rho-\tau)  m^2 - \frac{\pi i(\rho-\tau)}{R^2} k^2 + \pi i (\rho+\tau)mk}	\nonumber \\&\hskip2cm \times e^{2\pi i l(mR+k/R) (\tau J_{x0,0}-J_{y0,0}) -\frac{2\pi il^2}{\rho-\tau} (\tau J_{x0,0}-J_{y0,0})^2}\nonumber\\&=  \frac{\sqrt{\rho-\tau}}{\sqrt{2i}R}  e^{   -\frac{2 \pi i }{\rho-\tau} (\tau J_{x0,0}-J_{y0,0})^2}\nonumber\\
&\hskip1cm \times\sum_{m,k}	e^{-2\pi il(mR+k/R)J_{y0,0}} q^{\frac14(mR+k/R)^2+(mR+k/R)l  J_{x0,0}}\tilde q^{\frac14(mR-k/R)^2} .
\end{align}
Putting everything together gives:
\begin{align} 
Z(\tau,\rho) &\sim	 \frac{1}{\eta(q)\eta(\tilde q)}e^{-S_{n.l.}}e^{\frac{i}{2\pi}\int {\cal J} \wedge    \tilde \star J } e^{  - \frac{2 \pi i }{\rho-\tau} (\tau J_{x0,0}-J_{y0,0})^2}\nonumber\\&\hskip1cm \times\sum_{m,k}	e^{-2\pi i(mR+k/R)J_{y0,0}} q^{\frac14(mR+k/R)^2+(mR+k/R) l J_{x0,0}}\tilde q^{\frac14(mR-k/R)^2} . \end{align}
However, we can further simply this by separating out the zero modes in  ${\cal J}\wedge \tilde\star J$
\begin{align}
	e^{\frac{i}{2\pi}\int {\cal J} \wedge    \tilde \star J }=& e^{\frac{i}{2\pi}\int{\cal J}_{o} \wedge    \tilde \star J_{o} } 
	e^{    \frac{2 \pi i l^2 }{\rho-\tau} (\tau J_{x0,0}-J_{y0,0})(\rho J_{x0,0}-J_{y0,0})} ,
\end{align}
and hence we find
\begin{align} \label{Pfunction}
Z(\tau,\rho) &\sim	 \frac{1}{\eta(q)\eta(\tilde q)}e^{-S_{n.l.}}e^{\frac{i}{2\pi}\int{\cal J}_{o} \wedge    \tilde \star J_{o} }  e^{   2 \pi i l^2 (\tau J_{x0,0}-J_{y0,0}) J_{x0,0}}\nonumber\\&\hskip1cm \times\sum_{m,k}	e^{-2\pi i l(mR+k/R)J_{y0,0}} q^{\frac14(mR+k/R)^2+(mR+k/R) l J_{x0,0}}\tilde q^{\frac14(mR-k/R)^2} . \end{align}
Note that the   non-zero modes of the sources only appear in the first two exponential terms. 
It is interesting to observe that the source's zero-modes only appear coupled to $\tau$ and not $\rho$. In 
the absence of sources we find more simply  
\begin{align}\label{simpleZ}
Z(\tau,\rho) &\sim	 \frac{1}{\eta(q)\eta(\tilde q)} \sum_{m,k}	 q^{\frac14(mR+k/R)^2 }\tilde q^{\frac14(mR-k/R)^2} ,\end{align}
which is of the form given in (\ref{Z}) that we discussed in the  introduction and also agrees with the 
partition function (\ref{partfun}) calculated by canonical quantisation (taking $\beta = 2\pi i l$).

\subsection{Convergence}

In the analysis above we have proceeded formally and assumed that the integrals we discussed are convergent where necessary.  Let us now consider more carefully the criteria for their convergence. In full generality, we have a four-complex dimensional space of parameters   $\rho,\tilde\rho, \tau,\tilde\tau $. To simplify our analysis in this subsection, let us consider the case in which $\tilde\rho=\rho^*$ and $\tilde\tau = \tau^*$ which reduces our parameters to a two-dimensional complex  subspace corresponding to the case where both $g$ and $\bar g$ are Riemannian. 

We recall that our calculation  consisted of two path integrals: one over $Q_+'$, which leads to an effective action involving $P$, and then we must perform a path integral over $P$. 
We start with looking at  when the integration over $Q'_+$ is valid. In particular, the  condition that $Q'_+$ integral is a standard Gaussian is that ${\rm Im}( {\det }(\bar e)M_{--})>0$. 
As discussed earlier, the integral remains divergent even in this case, but can be regularised.
If we were to consider the case of Lorentzian-signature metrics $\bar g$ and $g$, where all metric components are real, then we would   have simply ${\rm Im}( {\det }(\bar e)M_{--})=0$ and hence the integral would not be Gaussian. Turning to the Euclidean context where $\rho$ and $\tau$ are complex we find
\begin{align}\label{bound0}
{\rm Im}(\det \bar e M_{--}) = e^{2\sigma}\frac{\rho_2}{|\rho-\tau^*|^2}	\left((\rho_1-\tau_1)^2 + \tau_2^2-\rho_2^2\right) ,
\end{align}
where $\rho = \rho_1+i\rho_2$ and $\tau = \tau_1+i\tau_2$. 
We take the conformal factor
      $e^{2\sigma}$ to be real and positive. If  $\rho_2<0$ then we must have 
 \begin{align}\label{bound1}
\rho_2  < -\sqrt{\tau_2^2 + (\rho_1-\tau_1)^2} .
 \end{align}
Alternatively for  $\rho_2>0$ we find 
 \begin{align}
\rho_2 <\sqrt{\tau_2^2 + (\rho_1-\tau_1)^2} .
 \end{align}
 Either way, there is a open set of values of $\rho$ and $\tau $ for which the $Q_+'$ integral is convergent.

Next, let us examine the convergence of the path integral  arising from the integration over $P$. Looking at $S_{eff}$, we see that we expect this requires that  $\sqrt{h}h^{\mu\nu}\partial_\mu P \partial_\nu P$  be positive definite. Note that this is a slightly weaker than the so-called  KSW  condition  \cite{Kontsevich:2021dmb,Witten:2021nzp} which requires that, in addition to the 0-form $P$,  the standard action for a 1-form   also converges. A computation shows  that the eigenvalues of ${\rm Re }(\sqrt{h}h^{\mu\nu})$ are
 \begin{align}
 	\lambda_\pm =\frac{1}{|\rho-\tau|^2}\left(X\pm\sqrt{ X^2 -Y }\right) ,
 \end{align}
where
\begin{align}\label{bound2}
X  &= (1+|\rho|^2)\tau_2-(1+|\tau|^2)\rho_2\nonumber\\ 	
Y &= 2\tau_2^2\rho_2^2-(\rho_2^2+\tau_2^2)\rho_2\tau_2-(\rho_1-\tau_1)^2\rho_2\tau_2  .\end{align}
Thus positivity of ${\rm Re }(\sqrt{h}h^{\mu\nu})$ requires that both $\lambda_+$ and $\lambda_-$ are positive and hence $X>0$ and $Y>0$. In full generality, this is a slightly complicated condition. However, we can restrict to the case   $\rho_2<0$ and $\tau_2>0$  so that $X,Y>0$ and, from (\ref{bound1}), we also require  
\begin{align}\label{bound4}
	\rho_2^2-\tau_2^2 >(\rho_1-\tau_1)^2 .
\end{align}

On the other hand, we can also consider the explicit path-integral  calculation over $P$ that we performed above for the case of a torus.  There we found that the integrals converge if ${\rm Re}(\varphi^2)<0$ (recall $\varphi = \frac12(\rho-\tau)$) which is equivalent to 
\begin{align}\label{bound3}
(\rho_2-\tau_2)^2 > (\rho_1-\tau_1)^2 .
\end{align} 
By writing $\rho_2 = -(1+\alpha)\tau_2$ one sees that (\ref{bound4}) implies (\ref{bound3}).
There was also a  condition   ${\rm Re} A >0$ in the Poisson re-summation of the zero-modes  which becomes simply $ \tau_2-\rho_2>0$. 
Lastly, we find that convergence of the sum over $m$ in the $\eta$-functions requires that $ \tau_2 >0$ and $ \rho_2 <0$.   Thus the explicit torus calculation requires $\tau_2>0$, $\rho_2<0$ in addition to  (\ref{bound4}). These  were referred to as the ``microscopic" conditions in \cite{BenettiGenolini:2025jwe} and we see that, as found there, they are stronger than the condition $X,Y>0$ and (\ref{bound0}), which allow for more general solutions.  

Finally, let us comment on the relation of our calculation to the calculation in \cite{Andriolo:2021gen} which took $\bar g =\eta$, corresponding to the choice $\rho=-\tilde\rho=1$ and hence $\tilde q=1$ here. This case is not included in our discussion   as we require $\rho_2<0$ for convergence. However,  for the non-zero modes the problem is simply an over-all physically irrelevant constant multiplicative factor.  The associated zeta-function regularisation is not sensitive to this  and the analysis of \cite{Andriolo:2021gen} found a single factor of $1/\eta(q)$ rather than $1/\eta(q)\eta(\tilde q)$.   When looking  at the zero-modes, the double sum we found here factorises into two single sums, one of which leads to a delta-function implying that $R^2=r_1/r_2$ is rational  and the remaining sum becomes a single-sum  theta-function. The divergent coefficient arising from  this delta-function was interpreted in \cite{Andriolo:2021gen} as arising from the unphysical modes that couple to $\bar g =\eta$ and not $g$.  In this way, the analysis of  \cite{Andriolo:2021gen} led to a partition function that is a single factor of $1/\eta(q)$ times a  theta-function of $\tau$. This is broadly consistent with  our result (\ref{simpleZ}) if we simply drop $1/\eta(\tilde q)=1/\eta(1)$   and $\sum \tilde q^{\tfrac14(mR-k/R)^2} =\sum 1$ as   constant but irrelevant divergent prefactors.  

\section{Modular Properties And Anomalies}

\subsection{Modular Properties}

Geometrically modular invariance arises from requiring invariance under large diffeomorphisms of the torus. However, it is well-known that the partition function for a single chiral boson is not modular invariant.  This is sometimes used as an argument why a single chiral boson cannot have a satisfactory action formulation as one would  expect the path integral to give a modular invariant expression on the torus.  Therefore it is of interest to examine what happens for Sen's action and the bi-metric action.

Before discussing our result and its generalisations, let us review the partition function obtained from the textbook canonical quantisation analysis on a spatial circle $x\cong x+2\pi$.  First, we consider a non-chiral, non-compact boson  where the mode expansion is $P = p_0 + Ex^0 + oscillators$. The partition function is 
\begin{align}
Z(\tau,\tau^*) & = \int dE  \sum_{oscillators} q^{L_0}(q^*)^{\tilde L_0}\nonumber\\
& \sim \frac{1}{\sqrt{\tau-\tau^*}}\frac{1}{\eta(q)\eta(q^*)}  .
\label{partfun1}
\end{align}
Here the factor of $({\tau-\tau^*})^{-1/2}$ arises from the integral over the continuous $E$ variable.  The $\eta$-functions arise from  sums over the  left and right moving oscillators. The zero-mode $p_0$ does not contribute to the energy -- this follows from time translational invariance -- and only contributes an irrelevant divergent factor to the partition function. This is well-known to be modular invariant.

In the case of a periodic scalar $P\cong P+2\pi R$, the mode expansion   is $P = p_0 + n x^0/Rl +mRx^1/l + oscillators$, and both  the winding and $E$ variables  become discrete: $m,n\in\mathbb Z$. Here find
\begin{align}
Z(\tau,\tau^*) & = \sum_{m,n}  \sum_{oscillators} q^{L_0}(q^*)^{\tilde L_0}\nonumber\\
& \sim \frac{1}{\eta(q)\eta(q^*)}\sum_{m,n} q^{\frac14(mR+n/R)^2 }(q^*)^{\frac14(mR-n/R)^2} .
\label{partfun2}
\end{align}
Again this is   modular invariant.
This partition function can be written as
\begin{align}
Z (\tau,\tau^*)  \sim \frac{1}{\eta(q)\eta(q^*)}\sum_{(p_L,p_R)\in \Gamma} q^{\frac12p_L^2 }(q^*)^{\frac12p_R^2} .
\label{partfun3}
\end{align}
where 
\begin{align}
p_L = \frac1{\sqrt{2}}(n/R+mR),\qquad p_R = \frac1{\sqrt{2}}(n/R-mR), 
\end{align}
and the sum is over all points $(p_L,p_R)$ in the Narain lattice $\Gamma$.
The modular invariance can be seen to follow from the fact that the Narain lattice is even and self-dual.

For rational values of   $R^2$, the Narain lattice decomposes into sublattices $\Gamma=\Gamma_L\oplus \Gamma_R$ with $p_L \in \Gamma_L$ and 
$p_R \in \Gamma_R$. Then  
\begin{align}
Z (\tau,\tau^*)  \sim \frac{1}{\eta(q)\eta(q^*)}\sum_{p_L \in \Gamma_L}\sum_{p_R \in \Gamma_R} q^{\frac12p_L^2 }(q^*)^{\frac12 p_R^2} .
\label{partfun4}
\end{align}
This then takes the form of a sum of holomorphically factorised terms:
\begin{align}
Z (\tau,\tau^*)  \sim \frac{1}{\eta(q)\eta(q^*)}\sum_a \vert f_a(q)\vert^2 ,
\label{partfun6}
\end{align}
for a finite set of generalised theta functions $f_a(q)$ (see e.g.\ \cite{Blumenhagen:2013fgp}).
In general the functions $f_a(q)$ transform into each other under modular transformations, so that all are needed for modular invariance in general.

For example, for the self-dual radius $R=1$ we find
\begin{align}
Z(\tau,\tau^*)  \sim \frac{1}{\eta(q)\eta(q^*)}\sum_{m,n} q^{\frac14(m+n)^2 }(q^*)^{\frac14(m-n)^2} .
\label{partfun5}
\end{align}
This can then be rewritten as a sum of holomorphically factorised terms:
\begin{align}
Z(\tau,\tau^*)   \sim \frac{1}{\eta(q)\eta(q^*)} (\vert \theta_3(\tau) \vert ^2+\vert \theta_{2} (\tau) \vert ^2) ,
\label{partfun61}
\end{align}
where the theta-functions are
\begin{align}
\theta_3(\tau)=  \sum_{m\in \mathbb{Z}} q^{ m^2} ,    \qquad \theta_{2}(\tau)=  \sum_{m\in \mathbb{Z}+\frac 1 2} q^{ m^2} .
\label{partfun88}
\end{align}
and these transform into each other under modular transformations.

For rational $R^2$ there are states with $p_R=0$,
 i.e.\  there are values of momentum and winding numbers $m,n$ for which $mR- n/R=0$. Restricting to these states defines a chiral boson theory with 
$p_L\in 
\Gamma_L$ 
and the partition function is
\begin{align}
Z (\tau)  \sim \frac{1}{\eta(q)}
\sum_{p_L \in \Gamma_L} q^{\frac12   p_L^2 }  .
\label{partfun7}
\end{align}
Here the self-duality condition implies that there is just one  set of left-moving  oscillators  and hence a single $\eta$-function and the partition function is not modular invariant. For example, for  the self-dual radius $R=1$  the states with $p_R=0$ have $m=n$ and $p_L=2m$ so that the partition function becomes
\begin{align}
Z(\tau) \sim 	\frac{1}{\eta(q)}\theta_3(q) , 
\label{partfun8}
\end{align}
 and this is not modular invariant.

Note that if  one tries to define a chiral boson theory when $R^2$ is not rational, the chiral constraint $mR- n/R=0$ implies $m,n=0$ so that the only states are oscillators with no momentum or winding
with $Z(\tau)   \sim 1/\eta(q)$. This is not modular invariant either. Furthermore, in both these cases we have had to impose self-duality by hand, as an additional constraint that does not follow from the action. 

We now look at the action of modular transformations on the   partition function $Z(\tau,\rho)$  we have obtained above using the Sen formulation, which does arise directly from a Lagrangian formulation with no additional constraints. 
In the case $\rho=\tau^*$ then $Z(\tau,\tau^*)$ agrees with the modular invariant partition function  given  in (\ref{partfun2}). 
This suggests that $Z(\tau,\rho)$ should be invariant under modular transformations in which both $\tau$ and $\rho$ transform. Indeed, it was shown in \cite{Hull:2023dgp} the action (\ref{action}) inherits two types of  diffeomorphism-like symmetries that act separately on $g$ and $\bar g$. However, the `true' diffeomorphisms, i.e. those that arise from coordinate transformations, correspond to the diagonal subgroup.  Thus, looking at the global diffeomorphisms of the torus that generate modular transformations, this implies that $Z(\tau,\rho)$ should only be invariant under modular transformations where which both $\tau$ and $\rho$ transform.  
That is, the full modular group is $SL(2,\mathbb{Z})_\tau \times SL(2,\mathbb{Z}) _\rho$ and $Z(\tau,\rho)$ should be invariant under the diagonal subgroup.
Let us now confirm this. 

 First let us consider the case where $P$ is not periodic and so does not admit any  winding modes. Here we find, setting the sources to zero for simplicity,  
\begin{align}
Z(\tau,\rho) \sim Z_{o}(\tau,\rho)  \sim \frac{1}{\sqrt{\tau-\rho}}\frac{1}{\eta(q)}\frac{1}{\eta(\tilde q)}  .	\label{partfun9}
\end{align} 
To examine its modular invariance we first note that  
 \begin{align}
\eta(\tau+1) = e^{\pi i/12}\eta(\tau) ,\qquad \eta(-1/\tau) = \sqrt{-i\tau}\eta(\tau)	 .
\end{align}
On the other hand we have defined $\eta(\rho)$ using $\tilde q = e^{-2\pi i \rho}$ which is valid in the lower half plane and so we have 
\begin{align}
\eta(\rho+1) = e^{-\pi i/12}\eta(\rho) ,\qquad \eta(-1/\rho) = \sqrt{i\rho}\eta(\rho)	 .
\end{align}
As a result, for a non-periodic scalar the partition function does not factorise nor does it have any nice properties under general modular transformations of $\tau $ and $\rho$. However, it  is invariant under diagonal modular transformations where  $\tau$ and $\rho$ both transform in the same way. 

Next we look at the case where $P$ is periodic: $P\cong P+2\pi R$. In addition to the non-zero mode contribution, we find the zero-mode contribution as given in (\ref{simpleZ}). 
For rational values of $R$, the partition function takes the form
\begin{align}
Z(\tau,\rho)   \sim \frac{1}{\eta(q)\eta(\tilde q)}\sum_a   f_a(q)  f_a(\tilde q)\label{partfun11} .
\end{align} 
For example, at the self-dual radius $R=1$ (\ref{partfun61}) is replaced by
\begin{align}
Z (\tau,\rho)  \sim \frac{1}{\eta(q)\eta(\tilde q)} ( \theta_3(\tau) \theta_3(\rho)+ \theta_{2} (\tau) \theta_{2 } (\rho)) .
\label{partfun611}
\end{align}

 Again it is easy to see that the partition function will not have any simple transformations properties under  $\tau\to\tau+1$ unless we also take $\rho\to\rho+1$ where it is invariant. Next taking both $\tau\to -1/\tau$ and $\rho\to -1/\rho$ we find that
\begin{align}
\frac{(n-\rho m)(n-m\tau)}{\rho-\tau}&\to 	\frac{(n+ m/\rho)(n+m/\tau)}{1/\tau-1/\rho}\nonumber\\
& = \frac{(n\rho+ m)(n\tau+m)}{\rho-\tau} .\label{partfun10}
\end{align}
This simply corresponds to swapping the sum $(m,n)\to (n,-m)$ in (\ref{Zzm}) and hence the zero-mode contribution is invariant under diagonal modular transformations.

Lastly we note that if we include the source zero-modes the diagonal modular invariance is preserved  they transform under $\tau\to \tau+1, \rho\to\rho+1$ as
\begin{align}
J_{x,0,0}\to  J_{x,0,0}\qquad  J_{y0,0}	\to J_{y0,0}+ J_{x0,0}
  ,
\end{align}
and under  $\tau \to -1/\tau$, $\rho\to -1/\rho$ as 
\begin{align} 
J_{x,0,0}  \to  
-J_{y,0,0}\qquad   J_{y,0,0}\to J_{x ,0,0} 	 .
\end{align}

\subsection{Gravitational Anomalies }

A single left-moving chiral scalar also has a 2-dimensional gravitational anomaly.  A consistent quantum theory can then only be found by combining it with other degrees of freedom. For example, the gravitational anomaly of  a left-moving chiral scalar can be cancelled by adding a right-moving chiral scalar or two right-moving real chiral Fermions.
However, such a theory of a right-moving chiral scalar plus  left-moving degrees of freedom is not modular invariant in general.

 The modular invariant partition function for a non-chiral scalar given above was
 \begin{align}
Z(q,q^*)   \sim \frac{1}{\eta(q)\eta(q^*)}\sum_{(p_L,p_R)\in \Gamma} q^{\frac12p_L^2 }(q^*)^{\frac12p_R^2}  ,
\label{partfun33}
\end{align}
and is a sum the sum  over all points $(p_L,p_R)$ in the Narain lattice $\Gamma$.
For rational values of   $R^2$, points in the Narain lattice with $p_L=0$ correspond to a right-moving chiral scalar and the ones with $p_R=0$ correspond to a left-moving chiral scalar. The partition function for a left and right chiral scalar given by restricting the sum to these states is not modular invariant.
For example, for the self-dual radius $R=1$ the restriction to chiral scalars is
\begin{align}
Z(q,q^*)   \sim \frac{1}{\eta(q)\eta(q^*)} \vert \theta_3(\tau) \vert ^2  ,
\label{partfun613}
\end{align}
which is not modular invariant. For the unrestricted scalar,  the modular invariant partition function  (\ref{partfun61}) has another term involving
$\theta_{2}$ coming from the points in the Narain lattice where both $p_L,p_R$ are non-zero.

As mentioned above, in the bi-metric chiral boson action, there are two gauge symmetries, one for which $g$ is the gauge field and one for which $\bar g$ is the gauge field, with the `true' diffeomorphisms arising as a diagonal subgroup of these \cite{Hull:2023dgp}. Anomalies arise for both these symmetries and these will be discussed more fully elsewhere (see \cite{Hull:2025bqo} and future work). 

The action  for a left-moving chiral scalar (\ref{action}) is invariant under both these symmetries, but the quantum theory has an anomaly in both.
This can be viewed as the physical scalar $A$ coupling to $g$ giving rise to an anomaly in the $g$-symmetry and the shadow sector $C$ coupling to $\bar g$ giving rise to an anomaly in the $\bar g$-symmetry.
However, if one adds a similar  action for a right-moving chiral scalar the anomalies in both symmetries are cancelled, as this introduces a right-moving physical scalar $\tilde A$ coupling to $g$ and a right-moving shadow scalar $\tilde C$ coupling to $\bar g$.
However, this theory is not  invariant under either the $\tau $ or $\rho$ modular transformations for reasons similar to those described above.
Such theories must then be coupled to other fields to achieve modular invariance. This will be explored in the following subsection.

Suppose now that two right-moving chiral fermions $\psi^a$  coupling to $g$ are added to the left-moving  chiral boson action (\ref{action}). This will cancel the $g$-symmetry anomaly but not the $\bar g$-symmetry anomaly. To do this a further two right-moving chiral fermions $\tilde \psi^a$ coupling to $\bar g$ are also needed. Thus the right-moving fermion theory also requires a shadow sector, if the anomaly in the b$\bar g$ symmetry is to be avoided. This too will be explored further in the next section.

\section{Chiral Bosons On A Torus, Anomalies And Modular invariance}

\subsection{Scalars On A Torus}

Consider first the standard theory of $n$ scalars taking values on an $n$-torus $T=\mathbb{R}^n/\Gamma$ for some lattice $\Gamma$, coupled to a world-sheet metric $g$. A constant $B$-field is incorporated by adding a topological term to the action that does not affect the equations of motion but which changes the definition of the canonical momentum.
 Then the momentum and winding vectors combine into a $2n$-vector $(p_L^i ,p_R^i)$ taking values in the $2n$-dimensional Narain lattice $\Lambda$, where $p_L^i $ is the momentum for left-movers and $p_R^i $ is the momentum for right-movers ($i=1,\dots, n$). The partition function is then a straightforward generalisation of (\ref{partfun4}):
\begin{align}\label{partfun3}
Z(\tau,\tau^*)   \sim \frac{1}{\eta(q)^n\eta(q^*)^n}\sum_{(p_L ,p_R)  \in \Lambda} q^{\frac12p_L^2 }(q^*)^{\frac12p_R^2} .
\end{align}
This is modular invariant as the $2n$-dimensional Narain lattice $\Lambda$ is even and self-dual. 

For special choices of the lattice and B-field (i.e.\ at special points in the Narain moduli space) 
 the Narain lattice decomposes into sublattices $\Lambda=\Gamma_L\oplus \Gamma_R$ with $p_L \in \Gamma_L$ and 
$p_R \in \Gamma_R$. These special lattices are the rational   lattices. For example, for $n=1$ the 2-dimensional lattice is rational when $R^2$ is rational.
For such a rational torus, the  partition function can be written as
\begin{align}
Z (\tau,\tau^*)  \sim \frac{1}{\eta(q)^n\eta(q^*)^n}\sum_{p_L \in \Gamma_L}\sum_{p_R \in \Gamma_R} q^{\frac12_L^2 }(q^*)^{\frac12 p_R^2} .
\label{partfun4}
\end{align}
Scalars on a rational torus define a rational conformal field theory, so that the partition function can be written in the form of a finite sum
\begin{equation}
Z(\tau,\tau^*)\sim
\sum _{a,\bar b} C_{a\bar b}\chi_a(\tau) \chi_{\bar b}(\tau^*)
\end{equation}
Here $C_{a\bar b}$ are constants, $a$ labels the left-moving primary fields and $\chi_a(\tau)$ is the character of the Verma module generated by the $a$'th primary field, while
 $\bar b$ labels the right-moving primary fields and $\chi_{\bar b}(\tau^*)$ is the character of the Verma module generated by the $\bar b$'th primary field.

On a rational torus, one can restrict to
 $n$ chiral bosons with $p_R^i=0$ and $p_L\in \Gamma'$ and the partition function is  
 \begin{align}
Z(\tau)   \sim \frac{1}{\eta(q)^n }\sum_{p_L   \in \Gamma_L} q^{\frac12 p_L^2 }  .
\end{align}
This is not modular invariant in general.

\subsection{The  Action for Chiral Bosons on a Torus Target Space}

The action (\ref{SenHact}) can be generalised to $n$ scalars $P^i$ and $n$ 1-forms $Q_{i+}$ (with $i=1,\dots n$) with a constant metric $G_{ij} $ 
and constant B-field $B_{ij}$. The action  is (including sources $J_{i\pm}$)
 \begin{align}
S &= -\frac{1}{2\pi}\int {\rm det}(\bar e)\Big(  -\partial_+P^i\partial_-P^j (G_{ij}+B_{ij})- 2 Q_{i+}\partial_-P^i-(Q_++J_+)_i(Q_++J_+)_j G^{ij}M_{--} \nonumber\\
&\hskip3
cm +2  G^{ij}( Q_{i+}J_{j-} + J_{i+}J_{j-}) \Big)	d^2x .
\end{align}
We will take $G_{ij} =\delta_{ij}$ and take the $P^i$ to be coordinates on a torus given by  $T=\mathbb{R}^n/\Gamma$  for some lattice $\Gamma$,
with $P^i\sim P^i+V^i$ for any $V^i\in \Gamma$.
By a change of coordinates, this could equivalently be written with scalars on a square lattice with $P^i\sim P^i+2\pi$ but with  a different metric, $G_{ij} \ne\delta_{ij}$.
As before, the term with $B_{ij}$ is a topological term that does not affect the equations of motion but which changes the definition of the canonical momentum.

The calculation of the partition function proceeds as before, with the result given by a sum over vectors in the Narain lattice $\Lambda$
\begin{align}
Z(\tau,\rho)   \sim \frac{1}{\eta(q)^n\eta(\tilde q)^n}\sum_{(p_L ,p_R)  \in \Lambda} q^{\frac12p_L^2 }(\tilde q)^{\frac12p_R^2} .
\end{align}
This agrees with (\ref{partfun3}) on replacing $q^* $ with $\tilde q$ and agrees with (\ref{simpleZ}) when $n=1$. 
At points in the Narain moduli space at which the Narain lattice is rational, the partition function takes the form
\begin{equation}
Z(\tau,\rho)\sim
\sum _{a,\bar b} C_{a\bar b}\chi_a(\tau) \chi_{\bar b}(\rho) .
\end{equation}

\subsection{Holomorphic Factorisation And Even Self-Dual Lattices}

For a standard bosonic CFT  the sum over oscillator modes gives a contribution to the partition function of the form
\begin{equation}
Z_o(q,q^*)\sim \frac{1}{\eta(q)^n\eta(q^*)^n} ,
\end{equation}
which factorises into a function of $\tau$ times a function of $\tau ^*$, but the inclusion of  momenta and winding from the zero modes in general gives a function of both 
 $\tau$ and  $\tau ^*$ that doesn't factorise.
 For the special tori for which the CFT is rational, the partition function is a finite sum of terms, each of which factorises into a function of $\tau$ and a function of $\tau ^*$.

For the Sen-type chiral boson action, the sum over oscillators gives a term which factorises into the product of a function of $\tau$ and a function of $\rho$, but now the momentum and winding contributions  give a function of both 
 $\tau$ and  $\rho$ in general. Just as before, in the rational case, this can be written as a finite sum of terms, each of which   factorises into a function of $\tau$ times  a function of $\rho$.

For certain lattices, there can be 
 points in the Narain moduli space for which there is a complete factorisation of the full partition function (including zero modes)
\begin{equation}
Z(\tau,\tau^*)\sim
F(\tau) G(\tau^*) ,
\end{equation}
for some functions $F,G$
and such points will have an important role here.
Such a complete factorisation arises at rational points in the Narain moduli space at which
$\Lambda=\Gamma_L\oplus \Gamma_R$ with $\Gamma_L= \Gamma_R=\Delta$ where $\Delta$ is an even self-dual lattice. This can only occur if  $n$ is a multiple of $8$. 
The partition function is then of the form \begin{equation}
Z(\tau,\tau^*)\sim F(\tau) F^*(\tau^*) .
\label{erhesdf} 
\end{equation}

For example, for $n=8$ there is only one even self-dual lattice and in this case  $\Delta$ is the root lattice of $E_8$.
In this case the partition factorises as (\ref{erhesdf})
where \cite{Lerche:1988np}\footnote{Note that \cite{Lerche:1988np} uses a convention where $\alpha'=2$ but we have $\alpha'=1$.}
\begin{equation}
F(\tau) = \frac{1}{\eta(q)^8}\Theta _{\Delta}(\tau) ,
\end{equation}
is the $\hat E_8$ character 
and
\begin{align}
\Theta _{\Delta}(\tau)=
\sum_{p  \in \Delta} q^{\frac12   p^2 }  ,
\label{partfunz7}
\end{align}
is the theta-function for the $E_8$ lattice. The $E_8$ theta function can   be written in terms of Jacobi theta functions as
\begin{align}
\Theta _{\Delta}(\tau)=\frac12 ( \theta  _2^8 (\tau)+\theta  _3^8 (\tau)+\theta  _4^8 (\tau)) .
\label{partfunz7s}
\end{align}

In these cases, the chiral partition function for  chiral bosons can be taken to be
\begin{equation}
Z(\tau )\sim
F(\tau) .  \end{equation}
Similarly for the corresponding Sen-type chiral boson theory the partition function is 
\begin{equation}
Z(\tau,\rho)\sim
F(\tau) F^*(\rho) .
\end{equation}

Next we consider a theory with $n$ chiral bosons with partition function $Z(\tau )\sim
F(\tau)  $ and embed it in a larger theory with other fields which couple to $g$ but not $\bar g$. For example we will consider the heterotic string in the next section. If the additional fields have a modular  partition function 
$f(\tau,\tau^*)$ then the total partition function 
\begin{equation}
Z(\tau,\tau^*)\sim f(\tau,\tau^*)F(\tau) ,
\label{FfPT}
\end{equation}
 is modular invariant.
If the $n$ chiral bosons are replaced by the  Sen-type chiral boson theory with partition function $Z(\tau,\rho)\sim
F(\tau) F^*(\rho)$, then the partition function for the whole system is of the form
\begin{equation}\label{partfun4}
Z(\tau,\rho)\sim f(\tau,\tau^*) F(\tau) F^*(\rho) ,
\end{equation}
which is still invariant under $\tau$ modular transformations. On the other hand, in the general case in which $Z(\tau,\rho)$ is a {\it sum} of terms that each factorise, then modular invariance would be hard, if not impossible,  to achieve.

Note that the partition function (\ref{partfun4}) 
may now have gravitational anomalies associated to $\bar g$. 
If $\bar g$ is treated as a fixed background field, then this may not be important. On the other hand, if we define a string-like theory in which we integrate over both  $\bar g$ as well as the physical metric $g$, then  $\rho$ modular invariance and no gravitational anomalies would be desirable.
This could be achieved by adding further matter coupling to  $\bar g$ but not $g$ with partition function $\tilde f (\rho,\rho^*)$ such that
$\tilde f (\rho,\rho^*)F^*(\rho)$ is modular invariant and anomaly free.  Then the total partition function
\begin{equation}
Z(\tau,\rho)\sim f(\tau,\tau^*) \tilde f (\rho,\rho^*)F(\tau) F^*(\rho) ,
\end{equation}
would be invariant under separate $\tau $ and $\rho$ modular transformations. Let us now consider the more specific example of the heterotic string.

\subsection{The Heterotic String}

The heterotic string can be formulated with 16 chiral bosons on $\mathbb{R}^{16}/\Delta$ where $\Delta$ is a 16-dimensional even self-dual lattice. There are precisely two such lattices: the root lattice of $E_8
\times E_8$ and the weight lattice of $Spin (32)/\mathbb{Z}_2$.
These both have the partition function
\begin{equation}
F(\tau) = \frac{1}{\eta(q)^{16}}\Theta _{\Delta}(\tau) ,
\end{equation}
where $\Theta _{\Delta}(\tau)$ is given by (\ref{partfunz7}).

The remaining degrees if freedom consist of 10 non-chiral uncompactified  scalars $X^\mu$,  10 right-moving fermions $\psi^\mu$,   the $b,c$ ghosts and the right-moving superghosts $\beta,\gamma$.
Then the total partition function is (\ref{FfPT}) where
\begin{equation}
f(\tau,\tau^*) =Z_b(\tau,\tau^*)Z_f(\tau,\tau^*) ,
\end{equation}
where 
\begin{equation}
Z_b(\tau,\tau^*)=\frac{1}{(\tau-\tau^*)^4}\frac{1}{\eta(q)^4\eta(q^*)^4}  ,
\end{equation}
is the partition function for 8 non-chiral uncompactified transverse scalars and
\begin{equation}
Z_f(\tau,\tau^*)=\frac{1}{(\tau-\tau^*)^4}( \theta  _3^4 (\tau^*)-\theta  _2^4 (\tau^*)-\theta  _4^4 (\tau^*)) ,
\end{equation}
is the partition function for 8 transverse right-moving fermions, with the 3 terms coming from the sum over spin structures.\footnote{This is the counting from the light-cone gauge formulation. In the covariant approach, $Z_b$ is the parition function for 10 scalars and the bosonic ghosts, while $Z_b$ is the parition function for 10 right-moving fermions and the superghosts.} 
For either choice of lattice $\Delta$, the resulting partition function
\begin{equation}
Z(\tau,\tau^*) =Z_b(\tau,\tau^*)Z_f(\tau,\tau^*)F(\tau)  ,
\end{equation}
is modular invariant.

Now we consider the same system of fields  $X^\mu,\psi^\mu,b,c,\beta,\gamma$ coupling to the physical metric $g$ together with the Sen-type action for fields $Q_i,P^i$ with $i=1,\dots, 16 $ coupled to both $g$ and $\bar g$, with the Narain lattice for the scalars $P^i$ chosen to be $\Lambda = \Delta\oplus\Delta$. The Sen-type action gives 16 left-moving scalars $A^i$ coupling to $g$ and 16 left-moving scalars $C^i$ coupling to $\bar g$.
The quantisation gives the partition function
\begin{equation}
Z(\tau,\tau^*,\rho) =Z_b(\tau,\tau^*)Z_f(\tau,\tau^*)F(\tau) F(\rho)  ,
\end{equation}
and this is invariant under the $\tau$ modular transformations, but not the $\rho$ modular transformations.
This construction cancels the gravitational anomaly for the $g$-gauge symmetries but not for the $\bar g$-gauge symmetries.

Invariance under the $\rho$ modular transformations can be achieved by adding a 2nd set of fields $\tilde X^\mu,\tilde \psi^\mu,\tilde b,\tilde c,\tilde\beta,\tilde\gamma$ coupling to $\bar g$ so that the partition function becomes
\begin{equation}
Z(\tau,\tau^*,\rho, \rho^*) =[Z_b(\tau,\tau^*)Z_f(\tau,\tau^*)F(\tau)] [Z_b(\rho,\rho^*)Z_f(\rho,\rho^*) 
F(\rho) ] ,
\end{equation}
which is invariant under both
$\tau $ and $\rho$ modular transformations.
Then the system consists of a physical heterotic string $X^\mu,\psi^\mu,b,c,\beta,\gamma$ coupling to the physical metric $g$ (and the physical world-sheet gravitino $\chi$) together with a shadow heterotic string $\tilde X^\mu,\tilde \psi^\mu,\tilde b,\tilde c,\tilde\beta,\tilde\gamma, C^i$ coupling to $\bar g$ 
(and  a shadow world-sheet gravitino $\tilde \chi$) with no coupling between the physical and shadow string theories, so that the shadow sector decouples and doesn't affect any of the physics of the physical sector.
Moreover, this construction cancels the gravitational anomaly for the $\bar g$-gauge symmetries,  so that both the gravitational anomaly for the $ g$-gauge symmetries and that
  for the $\bar g$-gauge symmetries are both anomaly-free.

Remarkably, a similar structure
was found in Sen's heterotic string field theory \cite{Sen:2015uaa}  and the new heterotic string field theory
\cite{Hull:2025mtb}, both of
which are formulated in terms of two string fields $\Psi, \tilde \Psi$ with $\Psi$ representing all the physical states of the heterotic string while $ \tilde \Psi$ gives rise to a second shadow  heterotic string with the same spectrum but which decouples from the physical string.

\section{Path Integral On A General Riemann Surface}

In this section we   extend the path integral discussion   to higher genus Riemann surfaces. To this end, we write the effective action as 
 \begin{align}\label{Action2}
 	\tilde S_{eff} &= 	-\frac{1}{2\pi}\int  \Big(\frac12 dP \wedge  \tilde \star d P -2{\cal J}\wedge  \tilde \star dP   +{\cal J} \wedge    \tilde \star J  \Big)	 \nonumber\\
 	& = -\frac{1}{2\pi}\int  \Big( -\frac12 P\wedge  \tilde\star \tilde \Delta P -2P\wedge d\tilde\star{\cal J}  +{\cal J} \wedge    \tilde \star J  \Big) .
 \end{align} 
 Here $\tilde \star$ is defined in analogy with the  Hodge dual:
 \begin{align}
 \tilde \star V_\mu = \sqrt{h}\epsilon_{\mu\nu}h^{\nu\lambda}V_\lambda	 ,
 \end{align}
 with   $\tilde \star^2=-1$ when acting on odd-forms,  $\tilde \star^2=1$ on even-forms, and
 \begin{align}
 \tilde \Delta P = 	 \tilde\star d \tilde\star dP ,
 \end{align}
 is the Laplacian associated to the complex metric $h$. 
We can see here that the path integral takes the form
\begin{align}
Z[J] = Z_{z.m.}Z_{o} e^{\frac{i}{2\pi}\int{\cal J} \wedge   \tilde\star  \tilde   J  }	 ,
\end{align}
where $Z_{z.m.}$ and $Z_{o}$ are the contributions arising from the zero-modes and non-zero-modes respectively. 

To continue, let us first consider the non-zero modes. Here we can start from the second form of the expression in (\ref{Action2}). We can shift the integral over $P$ to $P'= P+B$ with $B$ determined by
 \begin{align}
	\tilde \Delta  B= -2\tilde\star d\tilde\star{\cal J} .
\end{align}
This is possible since, for the non-zero modes, $\tilde \Delta $ is invertible allowing us to obtain $B$. This removes the linear term in $P$ from the action. However, the price of doing this is to introduce a non-local term in $S_{eff}$ of the form 
\begin{align}
S_{n.l.} = -\frac{1}{2\pi}\int dB\wedge \tilde\star {\cal J}	 .
\end{align}
Thus  we find
 \begin{align} 
 	 Z
 &= 	Z_{z.m} \int [dP']_{o} e^{-\frac{1}{4\pi}\int  P'\wedge    \tilde \star \tilde \Delta  P' } e^{\frac{1}{2\pi}\int {\cal J} \wedge    \tilde \star J  }	e^{-S_{n.l}} \nonumber\\
 &\sim\frac{1 }{\sqrt{{\rm det}'(\tilde \Delta )}} Z_{z.m} 	e^{-S_{n.l}}   e^{\frac{1}{2\pi}\int {\cal J} \wedge    \tilde \star J  }	 ,
 \end{align}
 where the determinant only includes the non-zero modes of  $\tilde\Delta$ and $Z_{z.m}$ represents the contributions from the zero-modes. 
 
 Choosing coordinates where  $\rho$ and $\tau $ constant the operator $\tilde \Delta $ factorizes: $\tilde \Delta  P=  4 \partial \tilde \partial   P$ where, as above, 
 \begin{align}
 \partial = -\frac{1}{\rho-\tau}\left(\partial_y - \rho \partial_x\right)\qquad \tilde \partial  = 	 \frac{1}{\rho-\tau}\left( \partial_y - \tau \partial_x\right) .
 \end{align}
 This means that the eigenvalues of the  Laplacian take the following form:
\begin{align}
\lambda_{\vec n}=		-\frac{4}{(\rho-\tau)^2}\lambda^{(\rho)}_{\vec n}\lambda^{(\tau)}_{\vec n} ,
\end{align}
where  $\vec n$ is a generic label for the eigenvalues and $\lambda_{\vec n}^{(\rho)}$ and $\lambda_{\vec n}^{(\tau)}$ only depend on $\rho$ and $\tau$ respectively. 
Formally the determinant is given by
\begin{align}
{\rm det}'(\tilde\Delta) = \prod_{{\vec n} }	\left(\frac{-2i}{\rho-\tau}\right)^2\lambda^{(\rho)}_{\vec n}\lambda^{(\tau)}_{\vec n} .
\end{align}
As before there will be a reality constraint on $P$ that relates the associated eigenfunctions and thereby restricts the product. 
 Using a zeta-function regularization this product is defined as 
\begin{align}
{\rm det}'(\tilde\Delta ) = e^{G'(0)}	 ,
\end{align}
where 
\begin{align}
G(s)  
& =  \left(\frac{-2i}{\rho-\tau}\right)^{2s}	\sum_{\vec n}(\lambda^{(\rho)}_{\vec n})^{s}(\lambda^{(\tau)}_{\vec n})^{s} \nonumber\\
& \equiv \left(\frac{2i}{\rho-\tau}\right)^{2s}\tilde G(s) .
\end{align}
It follows that
\begin{align}
G'(0) & = 2\ln\left(\frac{-2i}{\rho-\tau}\right)	\tilde G(0) + \sum_{\vec n}\ln  \lambda^{(\rho)}_{\vec n} + \sum_{\vec n}\ln \lambda^{(\tau)}_{\vec n} ,
\end{align}
and hence the determinant takes the form
\begin{align}
{\rm det}'(\tilde \Delta ) = 	\left(\frac{2i}{\rho-\tau}\right)^{2\tilde G(0)}  D_+(\rho) D_-(\tau) .
\end{align}
In this way the path integral on a generic Riemann surface  can be written as 
\begin{align} 
 	 Z  &\sim\left(\frac{-2i}{\rho-\tau}\right)^{- 2\tilde G(0)}\frac{1}{\sqrt{ D_+(\rho) D_-(\tau)}}  Z_{z.m}	e^{-S_{n.l}} e^{\frac{1}{2\pi}\int {\cal J} \wedge    \tilde \star J }	 .
 \end{align}

 To evaluate $Z_{z.m.}$ we recall that the zero-modes satisfy $\tilde\Delta P=0$.  To obtain these we note that $dP$ is a one-form and hence can be written as $dP= \alpha_++\alpha_-$ where $\tilde\star \alpha_\pm=\pm i\alpha_\pm$ and $d\alpha_++d\alpha_-=0$. The condition $\tilde\Delta P=0$ now imposes that $id\alpha_+-id\alpha_-=0$. Thus we find that $d\alpha_+=d\alpha_-=0$ and $d\tilde\star \alpha_+=d\tilde \star\alpha_-=0$. In the case of a real Euclidean metric on a Riemann surface these correspond to the holomorhic and anti-holomorphic one-forms respectively and $\alpha_+ = (\alpha_-)^*$.  
 
 Furthermore, these can have non-zero periods over the 1-cycles. This in turn means that $P$ is not globally defined but is allowed to be multi-valued although the periods are restricted, for example to be integer.  This means that the integration by parts that we used above is not valid. 
 
 To continue, we write $\alpha_+ = \sum_i\hat m_i\alpha^i_+$ and $\alpha_+ =\sum_i\hat n_i\alpha_-^i$ where the $i$ index labels 1-cycles of the Riemann surface.
Using the Riemann bi-linear identity we find, starting from the first line of (\ref{Action2}),
\begin{align}
Z_{z.m.} &= 	{\rm exp}\left(	\frac{1}{4\pi}\sum_{\hat m^i,\hat n^j}\int  \Big( i(\hat m_i\alpha^i_++\hat n_i\alpha^i_-)\wedge(\hat m_j\alpha^j_+-\hat n_j\alpha^j_-) -2i{\cal J}\wedge (\hat m_i\alpha^i_+-\hat n_i\alpha^i_-)    \Big)	\right)\nonumber\\
&={\rm exp}\Bigg(	-\frac{i}{2\pi}\sum_{\hat m^i,\hat n^j}\hat m_i\hat n_j\sum_k\Big(\oint_{A^k}\alpha_+^i\oint_{B_k} \alpha_-^j -\oint_{B_k}\alpha_+^i\oint_{A^k} \alpha_-^j\Big)\nonumber\\ &\qquad\qquad  -\frac{i}{2\pi}\sum_{\hat m_i }\hat m_i\sum_k\Big(\oint_{A^k}{\cal J}\oint_{B_k} \alpha_+^i -\oint_{B_k}{\cal J}\oint_{A^k} \alpha_+^i\Big)    \nonumber\\ &\qquad\qquad + \frac{i}{2\pi}\sum_{\hat n_i }\hat n_i\sum_k\Big(\oint_{A^k}{\cal J}\oint_{B_k} \alpha_-^i -\oint_{B_k}{\cal J}\oint_{A^k} \alpha_-^i\Big)  	\Bigg) .
\end{align}
To clean this up, we choose a basis where
\begin{align}
\oint_{A^k}\alpha_+^i & = \delta^i_k\nonumber\\
\oint_{B_k}\alpha_+^i & = {\cal T}^{ik}	 ,
\end{align}
and
\begin{align}
\oint_{A^k}\alpha_-^i & = \delta^i_k\nonumber\\
\oint_{B_k}\alpha_-^i & = \tilde {\cal T}^{ik}	 ,
\end{align}
so that, using a  vector notation, 
\begin{align}
Z_{z.m.}   
={\rm exp}\Bigg(	-\frac{i}{2\pi}\sum_{\hat{\bf n},\hat{\bf m}} \hat {\bf m}\cdot( \tilde {\cal T}^T-{\cal T})\hat{\bf n} -&\frac{i}{2\pi} \sum_{ \hat{\bf m}} \hat {\bf m}\cdot \left(   {\cal T} \oint_{\bf A}{\cal J}  -    \oint_{\bf B}{\cal J}\right)\nonumber\\
&+\frac{i}{2\pi}\sum_{\hat{\bf n} }  \hat {\bf n}\cdot \left(  \tilde  {\cal T} \oint_{\bf A}{\cal J}  -    \oint_{\bf B}{\cal J}\right)\Bigg) .
\end{align}
However, we would like to normalise the periods to
\begin{align}
\oint_{A^k} dP =2\pi R m_k\qquad 	\oint_{B_k} dP = 2\pi R n^k  ,
\end{align}
with $m_k,n^k\in {\mathbb Z}$ and $R$ a fixed constant. With our normalization we find\begin{align}
\hat {\bf m} =  -2\pi R ({\bf m}\tilde  {\cal T}    -{\bf n})({\cal T}-\tilde {\cal T})^{-1}	\qquad \hat {\bf n}=2\pi R ({\bf m} {\cal T}   -{\bf n})({\cal T}-\tilde {\cal T})^{-1}\ 
\end{align}
so that  $Z_{z.m.}$ becomes \begin{align}
Z_{z.m.}  
&={\rm exp}\Bigg(	- {2 \pi i R^2} \sum_{{\bf m},{\bf n}}(   {\bf m}\tilde{\cal T} -{\bf n})({\cal T}-\tilde {\cal T})^{-1}\cdot( {\cal T}-\tilde {\cal T}^T) ( {\bf m} {\cal T}   -{\bf n})({\cal T}-\tilde {\cal T})^{-1}\nonumber\\ &\qquad\qquad  +{2iR} ( {\bf m}\tilde{\cal T}  -{\bf n})({\cal T}-\tilde {\cal T})^{-1}\cdot \Big( \frac12({\cal T}+  \tilde {\cal T})\oint_{{\bf A}}{\cal J}-\oint_{{\bf B}}{\cal J} \Big)    	\Bigg) .
\end{align}

\section{Discussion and Conclusions}

In this paper we have studied the bi-metric generalisation \cite{Hull:2023dgp} of the  Sen formulation \cite{Sen:2015nph, Sen:2019qit} applied to chiral bosons on a Riemann surface. Analytically continuing these two metrics to Riemannian ones, we showed that the bi-metric  action reduces to that of a non-chiral boson coupled to a complex metric. We were then able to compute the partition function and gave an explicit expression for the case in which the 2d-space is a torus.   

Let us review our calculation. Formally, the Sen action is defined in terms of two Lorentzian metrics $\bar g$ and $g$ which we have parameterised by four independent real functions $\rho,\tilde\rho,\tau,\tilde \tau$ with $\rho\ne \tilde\rho$ and $\tau\ne\tilde \tau$ (as well as conformal factors that drop out of our calculations). As such, the action is real and the partition function, as defined by the path-integral in (\ref{Zdef}), is purely oscillatory and ill-defined. However, we can analytically continue $\rho,\tilde\rho,\tau,\tilde \tau$ to the complex plane where the action becomes complex and, at least for a certain range of $\rho,\tilde\rho,\tau,\tilde \tau$, it is damped allowing us to evaluate the partition function. In particular, this includes the  regime   $\tilde\rho =\rho^*$ and $\tilde\tau = \tau^*$, where  $\bar g$ and $g$ are Riemannian metrics, provided we take $\tau_2>0$ and $\rho_2<0$ and sufficiently large in magnitude.

 

 We also discussed modular properties of our partition function and showed that in general it is only invariant under transformations where the two complex structures transform in the same way. This is expected since only these transformations arise from coordinate transformations acting on the torus. In particular the zero mode contributions to the partition function do  not factorise.  However, for applications to the heterotic string theory we are interested in partition functions where $\tau$ and $\rho$ can be independently transformed. We showed that by considering $n$ chiral scalars, with $n$ a multiple of 8,  the corresponding partition function arising from a Narain Lattice can factorise into the simple form\begin{equation} 
Z(\tau,\rho)= F (\tau) \tilde F (\rho) \  ,
\end{equation}
and hence one can construct theories where the modular transformations of $\tau$ and $\rho$ can be made independent and moreover such that the theory is invariant under two sets of modular transformations. We showed how to construct a consistent worldsheet heterotic string theory by including anomaly cancelling matter fields in both the physical and shadow sectors.

 We have also  seen that in order to have a well-defined partition function we needed to have  $M \ne 0$ to provide the damping in the integrals.  
 Specifically, for a given physical metric $g$, we chose the metric $\bar g$ so that $M \ne 0$.
 If we were to try to  directly consider the case $M =0$, then the action would be simply
\begin{align}
	S = \int   Q'\wedge dP    ,
\end{align} 
where  $ Q'=  2 Q + \tfrac12 P+\tfrac12\bar \star dP  $. In two dimensions, this is a well-known conformal field theory; the holomorphic $\beta\gamma$   system,  with $c=2$. However, it generalises to higher dimensions where it resembles a topological $B\wedge F$  theory except that there is now a metric dependent constraint $Q'=\bar\star Q'$. We have examined this theory separately in \cite{Hull:2025bqo}.

\section*{Acknowledgements}

 We would like to thank Ashoke Sen for helpful discussions. The research of CH was supported by   the STFC Consolidated Grant    ST/X000575/1. N.L. is supported by the STFC grant ST/X000753/1.

\appendix
\section{Appendix: Conventions}\label{2dims}
 
In this paper we use two independent metrics $\bar g$ and $g$ and their associated Hodge-star operators are $\bar \star$ and $\star$. Furthermore, to avoid any confusion we denote complex conjugation by ${}^*$.
 
We adopt the following convention: the zwiebein indices are labelled as $a,b= +,-$. In particular we never use $+$ or $-$ for world indices. For example
\begin{align}
V_\mu &= \bar e_\mu{}^a V_a \\
&=	\bar e_\mu{}^+V_++ \bar e_\mu{}^-V_- ,
\end{align}
 and
 \begin{align}
 \partial_+ =\bar e_+{}^\mu\partial_\mu	\qquad \partial_- =\bar e_-{}^\mu\partial_\mu	 ,
 \end{align}
which in general are not  derivatives with respect to a pair of coordinates $x^+$ and $x^-$.  

Next we choose a convention where $\epsilon_{\mu\nu}$ is defined to be anti-symmetric with $\epsilon_{01}=1$. Similarly  $\epsilon^{\mu\nu}$ is anti-symmetric with $\epsilon^{01}=-1$. This implies that $\epsilon_{+-}=-\epsilon^{+-}=1$.

Given a metric, say $\bar g$, and one-form $V = V_\mu dx^\mu$ we define
\begin{align}
\bar\star V = \sqrt{-\bar g} \epsilon_{\mu\nu}\bar g^{\nu\lambda} V_\lambda	dx^\mu .
\end{align}
It follows that $\bar\star^2 V = V$ for any one-form $V$ and for any metric $\bar g$, including complex ones.

The benefit of the zwiebein basis is that
\begin{align}
\bar\star Q_a = \epsilon_{ab}\eta^{bc}Q_c	 ,
\end{align}
{\it i.e.}
\begin{align}
\bar\star Q_\pm = \pm Q_\pm	 ,
\end{align}
and hence a condition such as $Q=\bar\star Q$ simply implies that $Q_-=0$.
 
Lastly for a two-form $\omega^{(2)}=\tfrac12\omega_{\mu\nu}dx^\mu\wedge dx^\nu$ we define
 \begin{align}
\int \omega^{(2)} = \frac12\int \omega^{(2)}dx^\mu\wedge dx^\nu  = -\frac12\int \omega^{(2)}_{\mu\nu}\epsilon^{\mu\nu} d^2x .
\end{align}

\end{document}